\documentclass[12pt, a4paper]{article}

\usepackage[
  margin=0.75in,
  headsep=10pt, 
]{geometry}

\usepackage{graphicx}
\usepackage{soul}
\usepackage{graphicx}
\usepackage{epstopdf, epsfig}
\usepackage{amsmath}
\usepackage[table]{xcolor}
\usepackage{subcaption}
\usepackage{soul}
\usepackage{makecell}
\usepackage{multirow}
\usepackage{breqn}
\usepackage{amssymb} 
\usepackage{booktabs}
\usepackage{dcolumn}
\usepackage{bm}
\usepackage[utf8]{inputenc}
\usepackage[T1]{fontenc}
\usepackage{mathptmx}
\usepackage{etoolbox}
\usepackage{hyperref}
\usepackage{subcaption}
\usepackage{ragged2e}
\usepackage[sort&compress]{natbib} 
\usepackage[utf8]{inputenc}
\usepackage[T1]{fontenc}
\usepackage{graphicx}
\usepackage{epstopdf, epsfig}
\usepackage{xcolor}
\usepackage[table]{xcolor}
\usepackage{soul}
\usepackage{makecell}
\usepackage{multirow}
\usepackage{booktabs}
\usepackage{dcolumn}

\usepackage{float}
\usepackage{placeins}

\newcommand{\edt}[1]{{\color{black}#1}} 

\usepackage{amsmath}   
\usepackage{amssymb}   
\usepackage[normalem]{ulem}

\newcommand{\soptitle}{How do flapping avian wings exhibit superior aerodynamic performance?}
\usepackage{xcolor}

\begin{document}

\begin{center}
\Large \bf{\soptitle}
\vspace{0.1in}
\end{center}

\begin{center}
{Dilip Thakur and Muhammad Saif Ullah Khalid$^{\star}$}
\vspace{0.1in}
\end{center}
\begin{center}
Nature-Inspired Engineering Research Lab (NIERL)\\ 
Department of Mechanical and Mechatronics Engineering\\
Lakehead University, Thunder Bay, ON P7B 5E1, Canada\\
\vspace{0.05in}
$^\star$\small{Corresponding Author, Email: mkhalid7@lakeheadu.ca}
\end{center} 

\begin{abstract}
This \edt{work} investigates the unsteady aerodynamic performance and vortex dynamics of avian-inspired flapping wings using \edt{our} in-house sharp-interface immersed-boundary solver, \textit{VorteXdyn}. A falcon-inspired \edt{body-}wing model based on NACA $4312$ profile is employed to examine aerodynamic force generation and vortex evolution during steady forward flapping flight at Reynolds numbers of 2,500, $5\times10^3$, and $10^4$ and Strouhal numbers of $0.18$, $0.225$, and $0.27$. The influence of \edt{physiologies} is examined using three wing configurations: a simplified wing without distinct feather structures, a wing incorporating feather-like structures \edt{(serrations)} along the trailing edge, and a geometrically detailed wing incorporating multiple feather layers consisting of primary, secondary, and median feathers \edt{over its span}. The aerodynamic performance of these configurations is quantified using the temporal \edt{profiles} and time-averaged characteristics of the lift and drag coefficients and the lift-to-drag ratio. The associated vortex dynamics are characterized through the formation and evolution of multiple leading-edge vortices (LEVs), their spanwise coherence, circulation, characteristic size, and persistence over the wings' surfaces. Particular emphasis is placed on the spanwise development of the LEVs from the root to the wingtip, their interactions with the tip vortices, and the resulting wake evolution during the downstroke. Our results demonstrate that increasing geometric fidelity modifies aerodynamic force production, the formation and evolution of multiple LEV structures, vortex-vortex and votex-wing interactions, and wake topology. These findings provide insight into the aerodynamic role of feather morphology in three-dimensional flapping flight for the aerodynamic design of efficient bio-inspired flapping-wing micro air vehicles.
\end{abstract}

\section{Introduction}
\label{sec:Intro}
Flapping-wing aerodynamics attracted significant attention from biologists and engineers in recent years\edt{,} because of its exceptional efficiency, agility, and flight stability \edt{\cite{sane2003aerodynamics}}. Unlike conventional fixed-wing aircraft\edt{s} that rely primarily on steady aerodynamic principles, birds, insects, and bats generate lift and thrust through highly unsteady \edt{oscillations of their wings} that produce complex vortex structures and unsteady wake. These vortex-dominated flows enable natural flyers to perform agile maneuvers, rapid accelerations, hovering, climbing, and efficient cruising over a broad range of Reynolds numbers ($\mbox{Re}$). The exceptional flight performance observed in nature motivate\edt{s} extensive investigations into the \edt{underlying physical} mechanisms governing flapping flight, while simultaneously inspiring the development of bio-inspired micro air vehicles (MAVs) and flapping-wing robots capable of operating in confined and complex \edt{environments \cite{liu2024vortices,shyy2008aerodynamics,ma2024designing}}.

Among various unsteady aerodynamic mechanisms responsible for \edt{generation of forces} in flapping flight, the leading-edge vortex (LEV) is consistently identified as one of the primary contributors to \edt{enhancement of lift}. During the translational phase of the wing\edt{'s} stroke, \edt{separation of flow} at the leading edge produces a coherent vortex that remains attached to the wing surface \edt{for} over a substantial portion of the flapping cycle. The low-pressure region associated with the attached LEV augments circulation around the wing, \edt{which helps produced significantly greater lift force than that} predicted by conventional \edt{theories of stead-state aerodynamics \cite{sane2003aerodynamics,shyy2008aerodynamics,dickinson1999wing,liu2024vortices,wang2014lift}}. In addition to \edt{formation of $\mbox{LEV}$}, \edt{wing-wake interactions}, rotational circulation, and other unsteady vortex mechanisms contribute significantly to the aerodynamic performance of flapping wings, particularly at low and moderate Reynolds numbers\edt{,} where viscous effects remain important \edt{\cite{wang2005dissecting,shyy2008aerodynamics}.} \edt{Numerous} experimental and \edt{computational} investigations \edt{provided evidence of} highly three-dimensional and dynamically evolving topology \edt{of LEVs over oscillating wings}. Experimental flow visualizations \edt{revealed} the existence of a dual-LEV system comprising a primary attached vortex and a secondary co-rotating vortex near the leading edge \edt{of a wing \cite{lu2006dual,lu2008three}}. The evolution of these \edt{vortical substructures} depends on \edt{the} Reynolds number, \edt{kinematics of wings}, and \edt{their localized} angle\edt{s-}of\edt{-}attack \edt{\cite{lehmann2007aerodynamic,shyy2007flapping,liu2009integrated}}. \edt{Nevertheless,} relatively simpler and \edt{more} coherent LEV structures \edt{were} observed at low Reynolds numbers\edt{,} and increasingly complex \edt{vortex dynamics} \edt{occurred} as the Reynolds number increases \edt{\cite{birch2004force}}. \edt{For $\mbox{Re}\sim10^4$} \edt{that is a} characteristic of \edt{small-scale} avian flight, the flow \edt{undergoes} complex \edt{transitions and interactions} between inertial and viscous effects. High-fidelity numerical studies \edt{demonstrated} that \edt{stability of vortices, topologies of wakes}, and aerodynamic forces \edt{were} strongly influenced by kinematics \edt{of oscillating wings}, \edt{their} aspect ratio\edt{s} and \edt{advanced ratio \cite{TIAN2013149}. Particularly}, spanwise flow, tip vortices, and trailing-edge vortices \edt{factor in} the formation, growth, stability, \edt{shedding,} and\edt{/or} eventual breakdown of the LEV \edt{due to the vortex-vortex interactions happening over the wings \cite{gonzalo2025quantitative}}. \edt{All these features and processes influence} instantaneous aerodynamic forces and wake \edt{dynamics} throughout the flapping cycle \edt{\cite{Harbig_Sheridan_Thompson_2014, cheng2014wake}.}



In addition to these unsteady aerodynamic mechanisms, \edt{wings of birds, or avian wings} are highly morphing structures that continuously modify their shape through folding, spanwise extension, \edt{deployment of feathers}, and \edt{twisting} to adapt to varying flight conditions. Such \edt{dynamical and morphing wings} enables birds to regulate aerodynamic loading while enhancing maneuverability and efficiency \edt{during flight \cite{wong2025wing,de2024bio}}. Inspired by these biological adaptations, actively controlled morphing wings \edt{shown} shown to improve thrust generation, propulsive efficiency, and overall aerodynamic performance by providing additional degrees of freedom that allow the wing\edt{'s} geometry to adapt to different operating conditions \edt{\cite{lin2025flapping}}. Localized passive flow-control mechanisms further contribute to the aerodynamic performance of \edt{avian} wings. Experimental studies \edt{demonstrated} that covert feathers automatically \edt{get deployed} under separated-flow conditions, \edt{which delayed} stall and \edt{improved} lift and drag by shifting the \edt{location of} flow separation \edt{over the wings} downstream \edt{that greatly helped keep the flow attached} over a larger portion of the wing \edt{\cite{johnston2012investigation}}. Beyond large-scale morphing, local geometric features of biological wings also influence aerodynamic performance. Computational investigations of dragonfly-inspired corrugated wing sections in gliding flight \edt{showed} that natural corrugations \edt{could} generate lift comparable to or greater than that of smooth airfoils while maintaining similar drag, with the improved performance attributed to recirculation zones formed within the corrugations \edt{to} reduce shear drag \edt{\cite{kim2009effects,levy2009simplified}.} Furthermore, \edt{previous} experimental studies \edt{also reported} that increasingly realistic geometries \edt{of flapping wings} \edt{produced} higher lift and improved lift-to-drag ratios than simplified models \edt{of these wings that clearly} highlighting aerodynamic advantages of biologically inspired \edt{morphologies \cite{altshuler2004aerodynamic}}. \edt{However,} most \edt{of the relevant work on avian flight in existing literature employed} simplified \edt{geometries of wings}, two-dimensional configurations \edt{\cite{saddal2026effects}}, or isolated morphological features that \edt{lacked intricate} structural complexity of natural wings of birds \cite{vargas2008computational}. \edt{Besides, recent experimental work on avian flight either focussed on \edt{overall body's dynamics and its stability \cite{martinez2026tuning, weston2026stability}, simpler geometric features of wings \cite{hui2021experimental}, or low-fidelity mathematical models for estimating the performance metrics \cite{song2024investigation} of the unsteady flight.}}
Consequently, the combined aerodynamic influence of real-like primary and secondary feather\edt{-based} morphologies \edt{of avian birds on the governing development and dynamics of $\mbox{LEVs}$, tip vortices, and overall wake as well as the consequent generation of} unsteady forces \edt{for} a range of flow conditions \edt{with more realistic oscillatory kinematics remained largely unexplored, despite its relevance and importance.} 

 
Therefore, our present work \edt{aims at addressing the following research questions related to forward flight: (1) how does the addition of more sophisticated and real-like morphological features on avian birds, such as serrations and feathers, influence vortex dynamics, vortex-vortex interactions, vortex-wing interactions, and unsteady forces? and (2) what variations in the governing physical mechanisms for flapping flight happen when flow conditions and kinematic profiles are changed? For this purpose, we employ high-fidelity computational simulations, using our in-house solver, \textit{VorteXdyn}, for flows around a real-like falcon to resolve instantaneous intricate flow structures around its flapping wings and their connection with unsteady aerodynamic forces. These two research objectives and adoption of the presently reported computational methodology clearly explain the novelty, relevance, and significance of this work.} \edt{Our results are expected to provide new insights from avian flight for designing next-generation} bio-inspired flapping-wing systems.


\section{Computational Methodology}
\label{sec:Num_Method}

\subsection{\edt{Geometry \& Kinematics}}

\begin{figure}[H]
    \centering
    \makebox[\textwidth][c]{%
        \includegraphics[width=0.8\textwidth]{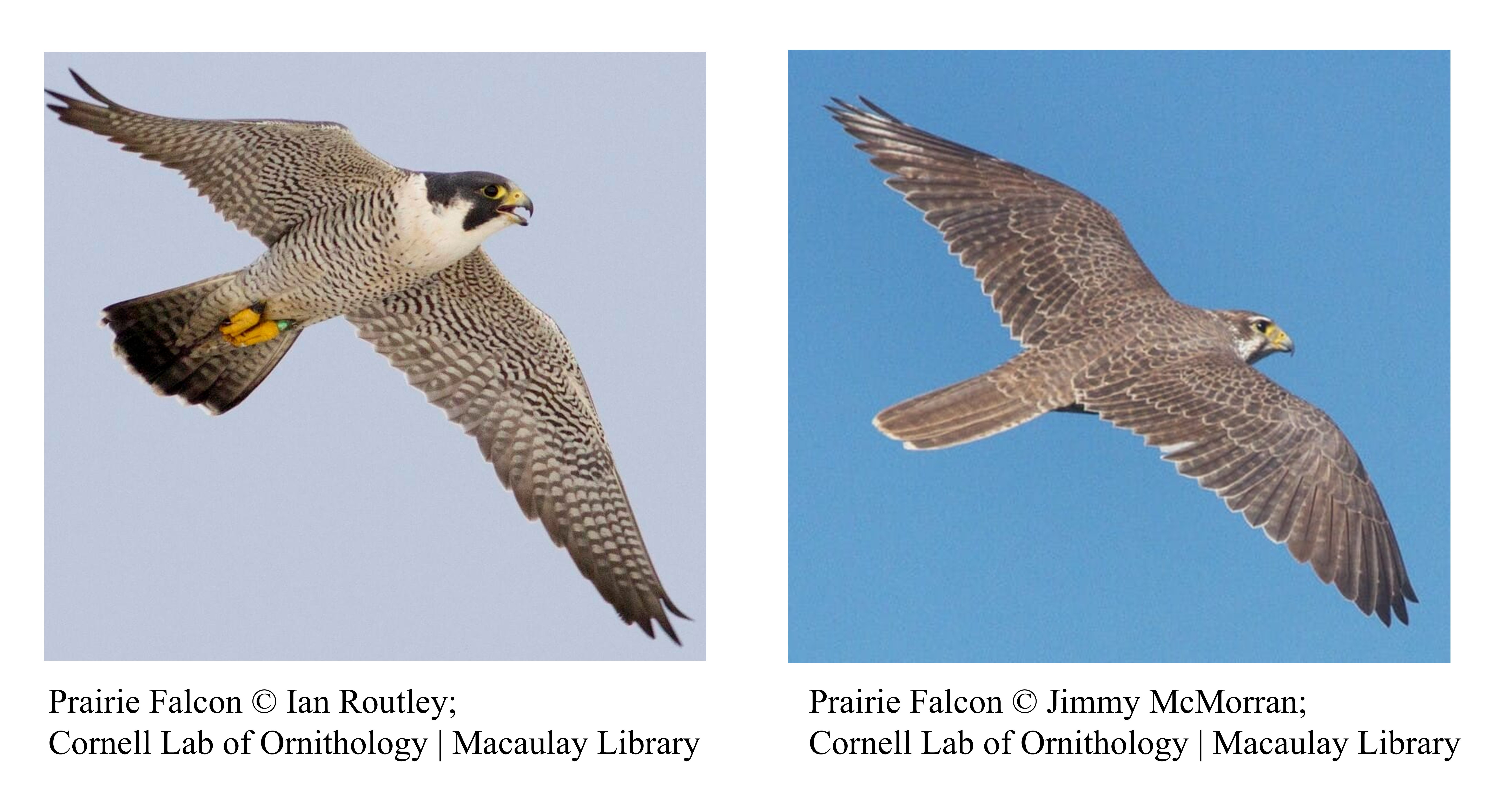}
    }
    \caption{Photographs of a Prairie Falcon (Falco mexicanus) in flight, illustrating the characteristic wing morphology and feather arrangement. Photographs: \edt{\copyright} Ian Routley (left) and \edt{\copyright} Jimmy McMorran (right), \edt{where the images are reprinted with permission} from \edt{the} Cornell Lab of Ornithology, Macaulay Library.}
    \label{fig:Prairie_Falcon}
\end{figure}

\edt{For our present research work, we reconstruct the main morphological features of the body and wings of Prairie falcon \textit{Falco mexicanus}, shown in Fig.~\ref{fig:Prairie_Falcon}} by maintaining \edt{its} characteristic body-length and \edt{span of wings}. \edt{It is carried out through computer-aided design (CAD) tools, including \textcolor{black}{SOLIDWORKS for Students 2025\textregistered{}} and \textcolor{black}{ANSYS Spaceclaim 2025 R2\textregistered{} }}. The geometric parameters are selected based on morphological data reported by the Cornell Lab of Ornithology \cite{CornellLab}, ensuring realistic representation of \edt{the} avian wing\edt{'s} structures. The cross-sections of \edt{the wings} are constructed using $\mbox{NACA4312}$ airfoil, which provides a representative cambered geometry similar \edt{to} a Pergrine falcon\edt{'s wings} during \edt{its} flapping motion. 

\begin{figure}[H]
    \centering
    \makebox[\textwidth][c]{%
        \includegraphics[width=1.0\textwidth]{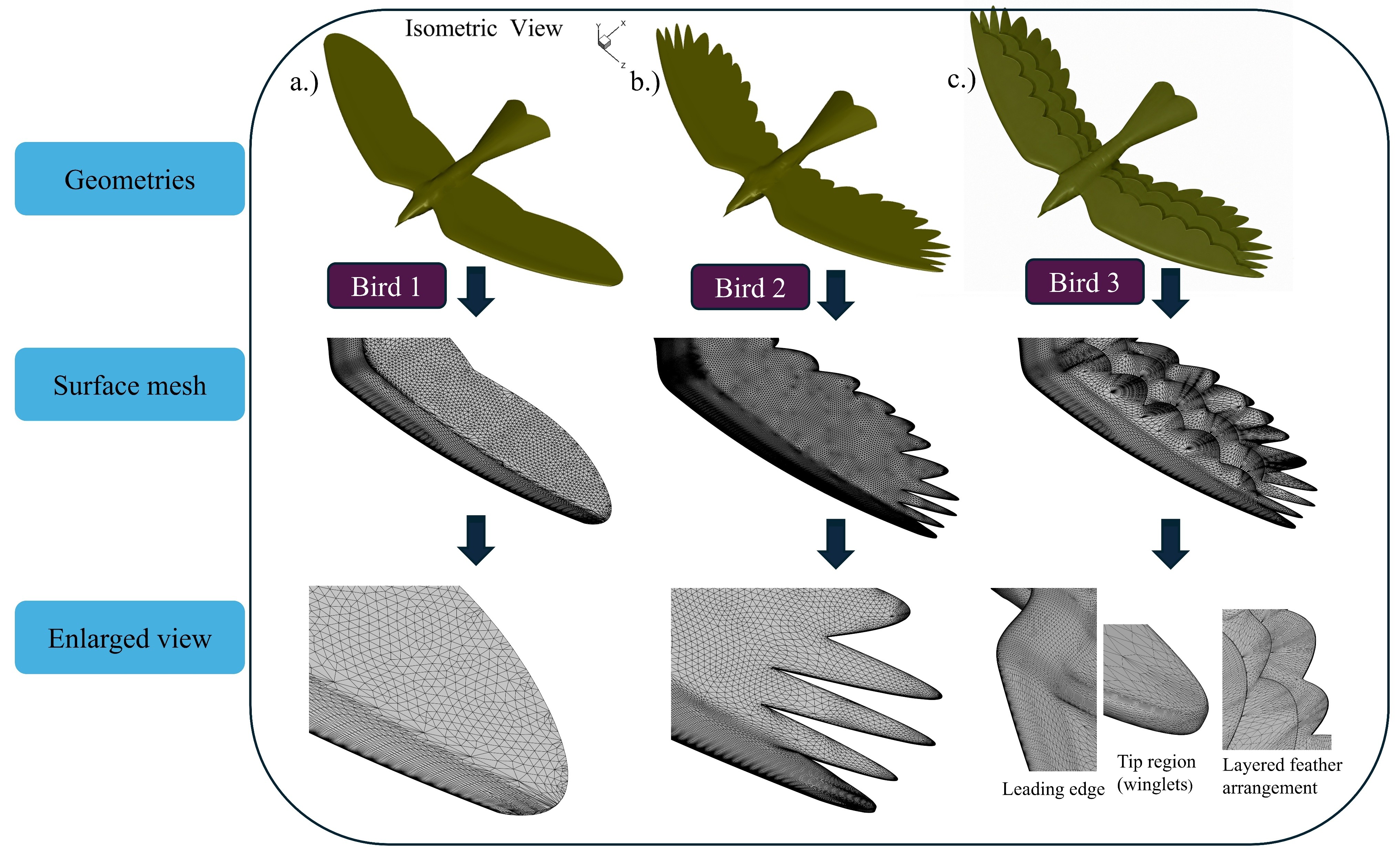}
    }
    \caption{\edt{Reconstructed morphologies of the bird and its wings in three different configurations along with the corresponding surface meshes}}
    \label{fig:Birds_with_Mesh}
\end{figure}

\edt{As illustrated in Fig.~\ref{fig:Birds_with_Mesh}, we consider} three \edt{geometric} configurations of wings with \edt{their} progressively increasing morphological complexity, which include \edt{(i)} a simplified wing, \edt{(ii)} a wing incorporating primary feathers, and \edt{(iii)} a wing incorporating both primary and secondary feathers. To isolate the aerodynamic effects of \edt{the} wing's morphology, \edt{our investigations focus} exclusively on the wings, \edt{whereas} the aerodynamic influence of the tail is not considered. The primary difference \edt{between} the three configurations lies in the detailed wing's morphology, particularly in the arrangement and distribution of layers \edt{over} the \edt{entire wings}.

All geometric \edt{quantities and characteristics} are nondimensionalized using the \edt{span of the wing ($S$)} as the reference length\edt{, setting the overall nondimensional span as $S^\ast=1$. T}he corresponding nondimensional body length is $L^\ast = 0.45S$. The planform area of wings based on the geometry is $0.1490$ for $\mbox{Bird~1}$, and $0.1540$ for $\mbox{Bird}~2$ and $\mbox{Bird}~3$. The latter two configurations share similar overall wing structures, with variations introduced in the geometric transition between the primary and secondary feathers \edt{over the wings}. \edt{During a complete flapping cycle, \textcolor{black}{the flapping motion is prescribed following the kinematic formulation presented by Beaumont et al. \cite{beaumont2025aerodynamic}. Rather than treating the wing as a completely rigid surface, spanwise geometric deformation is introduced here to reproduce a more realistic \edt{motion of an avian} wing when viewed from the frontal plane. Therefore,} the overall motion of the wings is constituted by two primary kinematic modes, including pitching\edt{,} and spanwise geometric deformation of the wing surface.} The instantaneous local pitching angle is expressed as\edt{:}

\begin{equation}
\theta(z,t) = \theta_0 \sin({2}{\pi}{f}{\edt{T}})\,\mathrm{sign}(z - z_{mid})\,W(z)
\end{equation}

\noindent where $\theta_0$ is the maximum pitching amplitude\edt{,} and $f$ \edt{denotes} the flapping frequency\edt{, $z$ shows the geometric coordinate along the span, and $T$ is the time}. The function $\mathrm{sign}(z - z_{mid})$ determines the direction of rotation and ensures antisymmetric motion between the left and right wings about the mid-span location $z_{mid}$. The spanwise variations \edt{during flapping are prescribed} through the weighting function $W(z)$, defined as\edt{:}

\begin{equation}
W(z) =
\begin{cases}
0, & |z - z_{mid}| \le d_{start} \\
\dfrac{|z - z_{mid}| - d_{start}}{d_{end} - d_{start}}, & d_{start} < |z - z_{mid}| < d_{end} \\
1, & |z - z_{mid}| \ge d_{end}
\end{cases}
\end{equation}

\noindent \edt{where $d$ indicates \textcolor{black}{the distance from the root to tip of a single wing}}. This function enforces a rigid region near the \edt{root of the wing} and gradually increases the amplitude \edt{for the motion} towards the \edt{tip of the wing.} The \edt{overall} geometric deformation\edt{s are} obtained by rotating each point about a spanwise hinge axis located at $z = z_{hinge}$. The transformed coordinates are given by:: 

\begin{align}
x' &= x \\
y' &= (z - z_{hinge}) \sin\theta + y \cos\theta \\
z' &= (z - z_{hinge}) \cos\theta - y \sin\theta + z_{hinge}
\end{align}

This formulation represents a purely kinematic model in which the wing deformation is prescribed and does not arise from structural dynamics, while still capturing the essential spanwise flexibility observed in flapping \edt{avian} wings. 

\subsection{Governing Equations for Flows and the Numerical Method}

\begin{figure}[H]
    \centering
    \makebox[\textwidth][c]{%
        \includegraphics[width=1.3\textwidth]{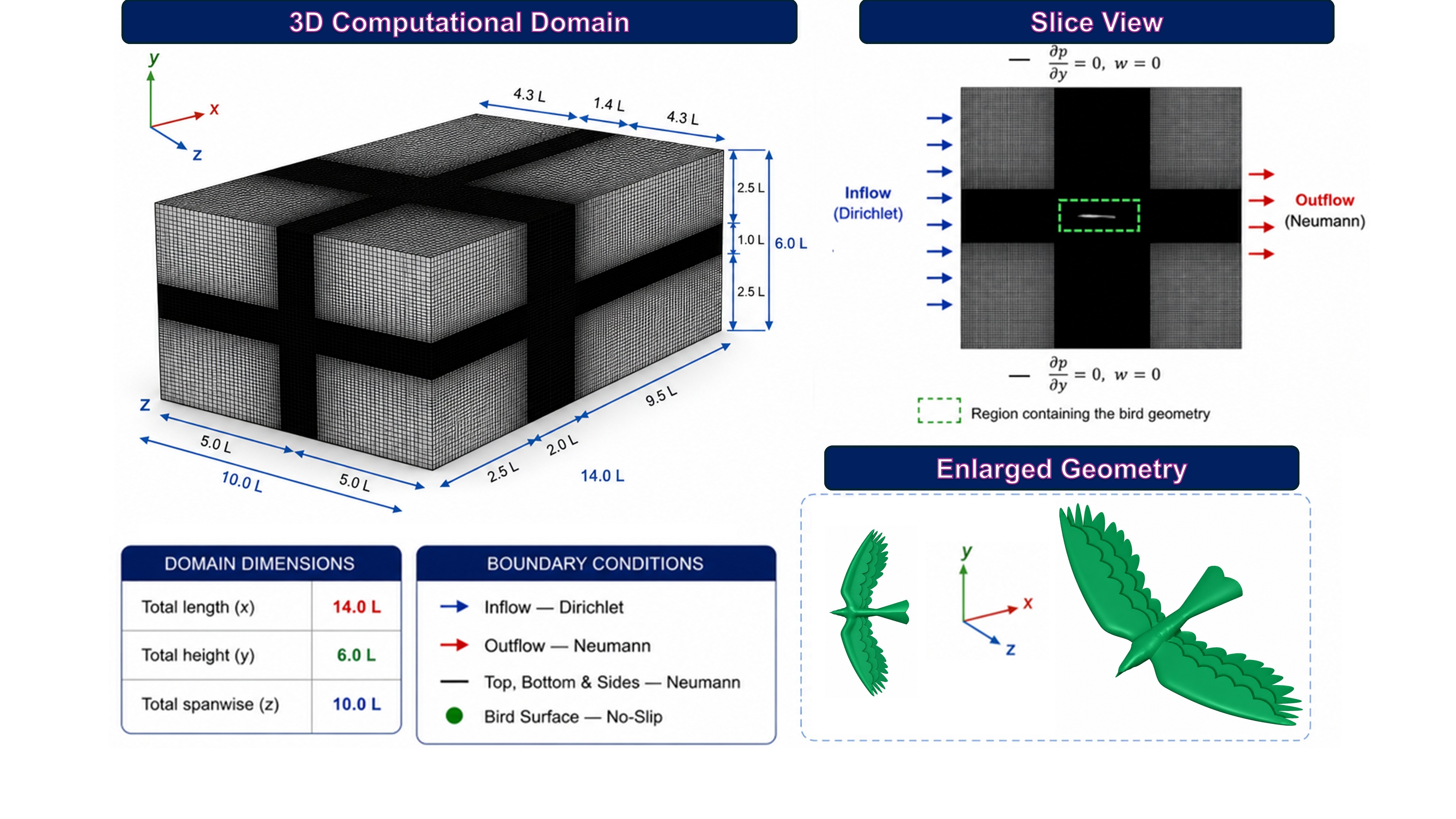}
    }
    \caption{Computational domain and the grid, showing local refinements around the body}
    \label{fig:grid}
\end{figure}

\edt{Dynamics of the} flow field is governed by the three-dimensional ($\mbox{3D}$), incompressible \edt{continuity and} Navier-Stokes equations in \edt{their} non-dimensional \edt{forms expressed below:}

\begin{equation}
\frac{\partial u_j}{\partial x_j} = 0
\end{equation}

\begin{equation}
\frac{\partial u_i}{\partial t} 
+ u_j \frac{\partial u_i}{\partial x_j}
= -\frac{1}{\rho} \frac{\partial p}{\partial x_i}
+ \frac{1}{Re} \frac{\partial^2 u_i}{\partial x_j \partial x_j}
+ f_b
\end{equation}

\noindent where $i,j = 1,2,3$, $x_i$ represents the Cartesian coordinates, $u_i$ denotes the Cartesian components of \edt{velocity of} the fluid, $p$ is pressure, and $\mbox{Re}$ is Reynolds number. \edt{In this formulation, $f_b$ is the discrete forcing term that enables a sharp representation of the immersed boundary \cite{farooq2025accurate, mittal2008versatile}. Also,} Reynolds number is defined as\edt{:}

\begin{equation}
\mbox{Re} = \frac{U_{\infty} c_m}{\nu}
\end{equation}

\noindent where $U_{\infty}$ is the free-stream velocity, $c_m$ \edt{represents} the mean chord length, and $\nu$ \edt{denotes} kinematic viscosity of the fluid. The mean chord length is calculated as\edt{:}

\begin{equation}
c_m = \frac{A_p}{b}
\end{equation}

\noindent where $A_p$ \edt{represents} the wing\edt{'s} planform area and \edt{its span, respectively}. The $\mbox{3D}$ computational domain is defined with dimensions of $14L \times 6L \times 10L$, where $L$ denotes \textcolor{black}{the overall length of a wing} in \edt{the} spanwise direction, taken as \edt{unit here}. This \edt{size of the} domain ensures adequate resolution of wake development and vortex dynamics while minimizing the influence of boundary conditions. \edt{The domain and the overall Cartesian mesh inside it is presented in Fig.~\ref{fig:grid}.}

The governing equations \edt{for flow dynamics} are solved using our in-house \edt{solver, \textit{VorteXdyn}, developed based on} sharp-interface immersed\edt{-}boundary method ($\mbox{IBM}$) \cite{farooq2025accurate}. The spatial derivatives are discretized using the second-order central difference scheme for the diffusion terms and the Quadratic Upstream Interpolation for Convective Kinematics (QUICK) scheme for the convective terms. The solver employs the fractional-step method for time integration, which ensures second-order temporal accuracy. A Dirichlet boundary condition is prescribed at the inlet, while Neumann boundary conditions are imposed at all other boundaries of the computational domain. The no-slip boundary condition on \edt{the bird's body} is enforced using the ghost cell-based immersed boundary approach \cite{farooq2025accurate}. \edt{Further details on the numerical methods used in this solver are available in Refs.\cite{farooq2025accurate, fardi2026fully}.}


\subsection{\edt{Validation \& Verification of the Computational Approach}}

\edt{First, we perform a grid-convergence test} to ensure that the numerical solution is independent of the spatial discretization \edt{of the domain}. The representative \edt{configuration of} $\mbox{Bird}~3$ at \edt{$\mbox{Re}=10^4$} \edt{at a Strouhal number ($\mbox{St}$) of} \textcolor{black}{0.27} is selected for \edt{this} analysis. Three \edt{different nonuniform structured grids}, referred to as coarse, medium, and fine grids, are \edt{employed with approximately $88~\mbox{million}$ grid points ($697 \times 309 \times 409$ nodes), $96~\mbox{million}$ grid points ($715 \times 317 \times 421$ nodes), and $102~\mbox{million}$ grid points ($733 \times 325 \times 427$ nodes), respectively. \textcolor{black}{The corresponding cell sizes in the fine region are $0.0036$, $0.0035$, and $0.0034$ for the coarse, medium, and fine grids, respectively.}} \edt{In order to ensure steady periodic state of aerodynamic forces, we run all simulations, including the actual ones reported latter in Section~\ref{sec:result}, for \textcolor{black}{$8-10$} {oscillation cycles of the flapping wings}. Figure~\ref{fig:CD_CL} \edt{plots comparisons between} \edt{temporal} histories \edt{of the drag coefficient ($C_D$) and the lift coefficient ($C_L$)} for one complete flapping cycle \edt{using the three grids}. The plots of both $C_D$ and $C_L$ \edt{from} the coarse, medium, and fine grids exhibit close agreement throughout the flapping cycle. Minor discrepancies are observed primarily near the peak aerodynamic loads, where the coarse-grid solution shows slightly larger deviations. The close overlap between the \edt{solutions from the} medium and fine grids indicates that further mesh refinement produces only small changes in the predicted aerodynamic forces. 




\begin{figure}[!ht]
    \centering
    \includegraphics[width=\textwidth]{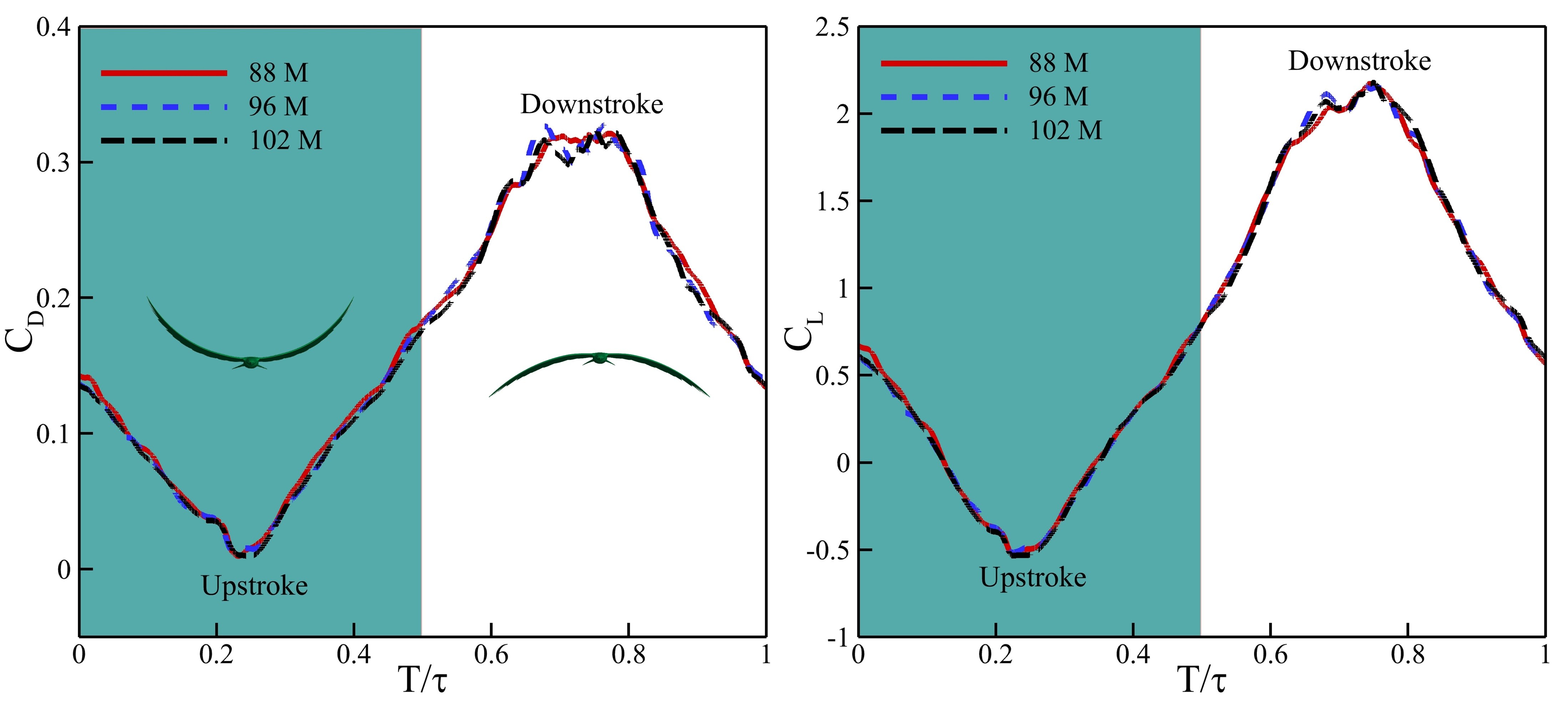}
   \caption{\edt{$C_L$ and $C_D$ over one complete flapping cycle for the three meshes used for the grid-convergence test}}
    \label{fig:CD_CL}
\end{figure}

\edt{Additionally,} as each simulations reaches its steady state solution in \textcolor{black}{$4$-$5$} {cycles}, \edt{we compute the time-averaged force coefficients, \textit{i.e.}, $\bar{C_L}$ and $\bar{C_D}$, for the last four flapping cycles. These coefficients and are reported} in Table~\ref{tab:grid_independence} along with their relative errors considering the solutions from the fine grid as the reference one}. The relative error is calculated as\edt{:}

\begin{equation}
\text{Relative error}~(\%) =
\frac{|\text{Grid value}-\text{Fine grid value}|}
{|\text{Fine grid value}|}\times 100.
\end{equation}



\begin{table}[H]
\centering
\caption{\edt{Results of the g}rid-independence \edt{tests}}
\label{tab:grid_independence}
\renewcommand{\arraystretch}{1.25}
\begin{tabular}{ccccc}
\hline
\textbf{Grid} & 
$\boldsymbol{\overline{C}_{L}}$ & 
\textbf{Error (\%)} & 
$\boldsymbol{\overline{C}_{D}}$ & 
\textbf{Error (\%)} \\
\hline
Coarse           & 0.8637 & 1.54 & 0.1749 & 1.27 \\
Medium           & 0.8616 & 1.29 & 0.1732 & 0.29 \\
Fine (reference) & 0.8506 & 0.00 & 0.1727 & 0.00 \\
\hline
\end{tabular}
\end{table}

\edt{Based on the temporal profiles of the aerodynamic force coefficients and their statistics, we choose the medium grid for our next simulations.}




\begin{table}[H]
    \centering
    \caption{\edt{Results for the time-step independence test}}
    \label{tab:time_independence}
    
    \renewcommand{\arraystretch}{1.25}
    
    \begin{tabular}{ccccc}
        \hline
        \textbf{Time Steps/Cycle} & 
        $\boldsymbol{\overline{C}_{L}}$ & 
        \textbf{Error (\%)} & 
        $\boldsymbol{\overline{C}_{D}}$ & 
        \textbf{Error (\%)} \\
        \hline
        3200 & 0.8658 & 0.743 & 0.1750 & 0.86 \\
        3600 & 0.8597 & 0.020 & 0.1747 & 0.73 \\
        4000 & 0.8595 & Reference & 0.1735 & Reference \\
        \hline
    \end{tabular}
\end{table}

\begin{figure}[!ht]
    \centering
    \includegraphics[width=\textwidth]{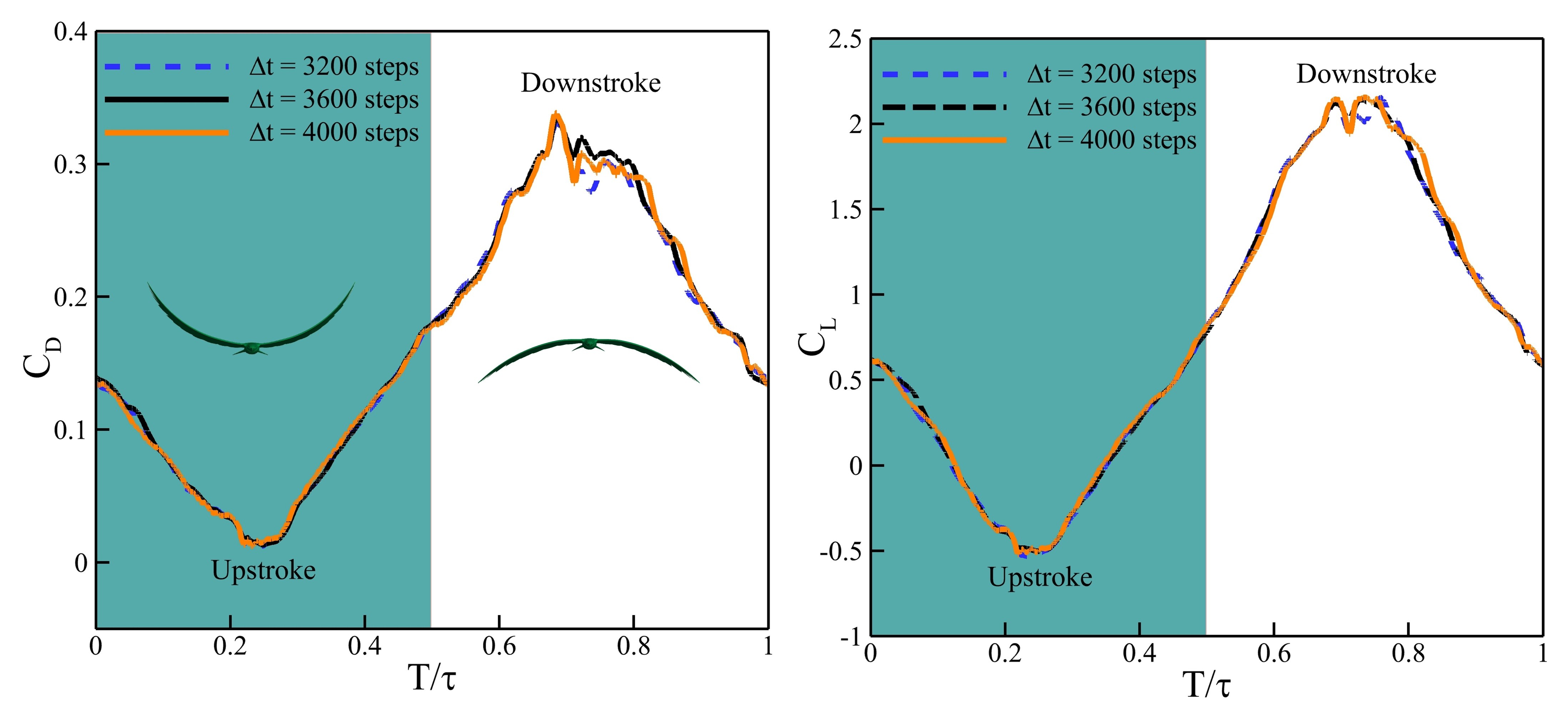}
   \caption{\edt{$C_L$ and $C_D$ over one complete flapping cycle for the three meshes used for the time-step independence test}}
    \label{fig:time_independence}
\end{figure}

\edt{Next, we conduct} a time-step independence study to assess the sensitivity of the numerical solution to temporal discretization. Three temporal resolutions of $3200$, $3600$, and $4000$ time steps per flapping cycle are considered while maintaining identical computational grid \edt{(the medium one)} and flow \edt{and kinematic} conditions \edt{described earlier. We compute $\bar{C_L}$ and $\bar{C_D}$ and evaluate} relative errors with respect to the finest temporal resolution\edt{,} as summarized in Table~\ref{tab:time_independence}. \edt{Comparative plots for unsteady $C_D$ and $C_L$ for one flapping cycle are provided in Fig.~\ref{fig:time_independence}. Both comparisons provide evidence of no significant variations in results, when the number of time steps is increased in the simulations. Therefore, we run our next simulations in this work using $3200$ time steps per flapping cycle.}



\begin{figure}[H]
\centering

\begin{subfigure}[b]{0.49\textwidth}
    \centering
    \includegraphics[width=\textwidth]{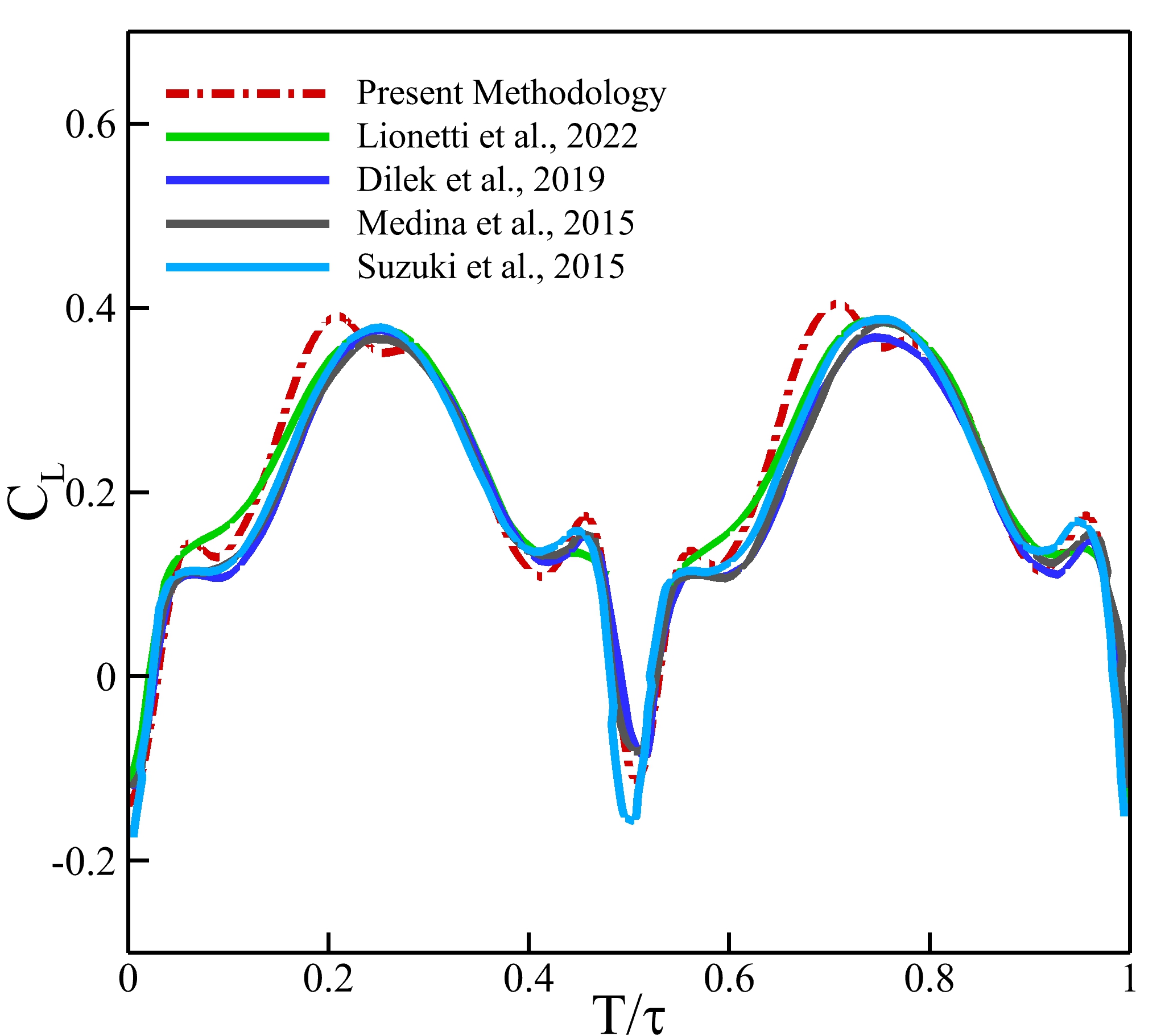}
    \caption{Lift coefficient ($C_L$).}
    \label{fig:Validation_CL}
\end{subfigure}
\hfill
\begin{subfigure}[b]{0.49\textwidth}
    \centering
    \includegraphics[width=\textwidth]{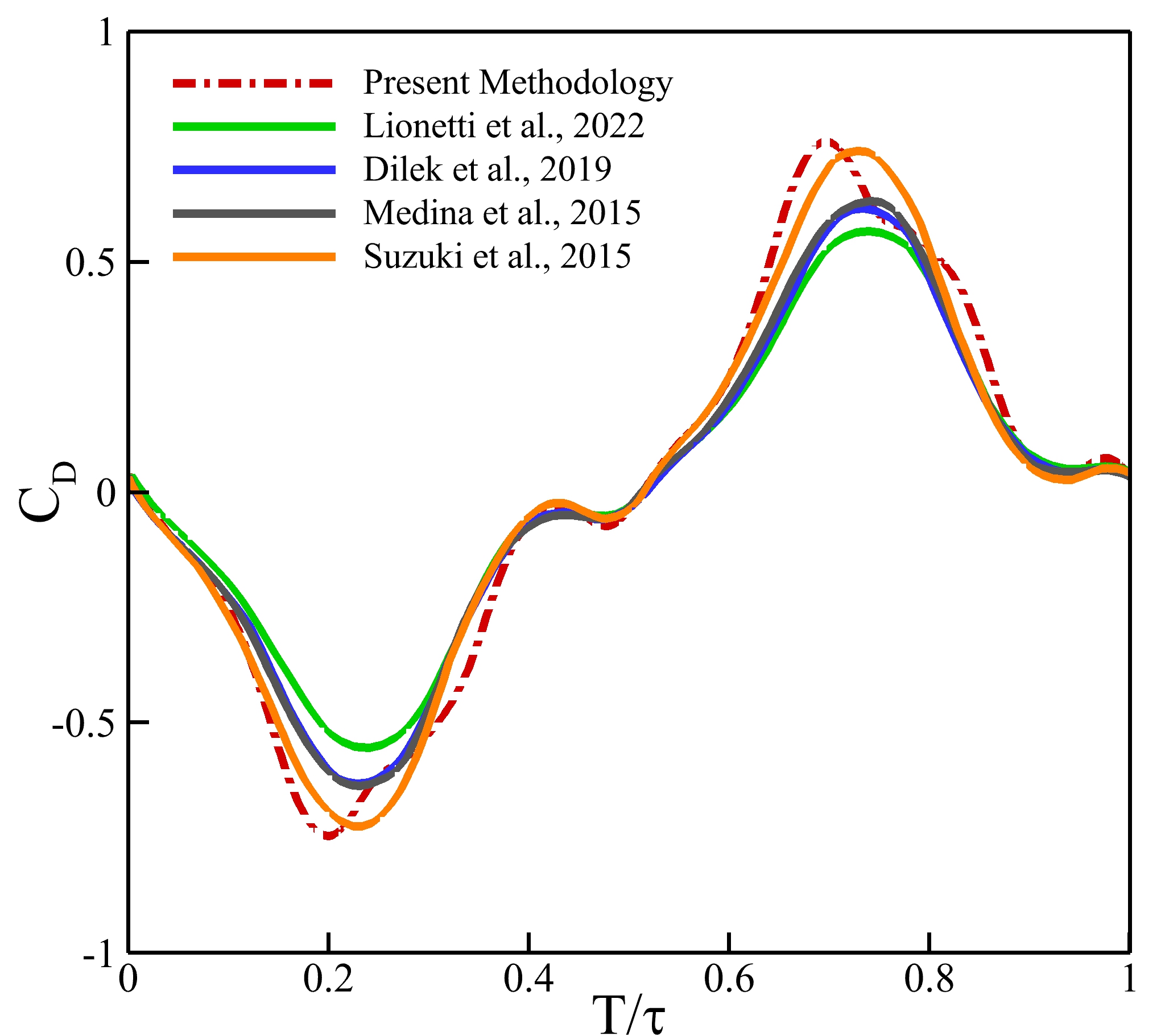}
    \caption{Drag coefficient ($C_D$).}
    \label{fig:Validation_CD}
\end{subfigure}

\caption{Comparison of the aerodynamic force coefficients \edt{computed} by the present solver with published numerical results for the benchmark rectangular flapping wing, \edt{originally employed in Suzuki et al.~\cite{Suzuki_Minami_Inamuro_2015}} }
\label{fig:Validation}
\end{figure}

To demonstrate the validity and accuracy of our solver, the benchmark flapping-wing configuration proposed by Suzuki et al. \cite{Suzuki_Minami_Inamuro_2015} is employed here. This \edt{case considers} a rectangular wing flapping in a horizontal stroke plane with \edt{no deviations}. \edt{This wing a spanwise length $L = 2c$} and a thickness of $w = 0.1c$ \edt{, where $c$ is its chord length}. \edt{Kinematics of the wing} are prescribed by \edt{(i)} a sinusoidal stroke motion \edt{mathematically expressed as} $\psi(t)=\psi_m\cos(2\pi ft)$, where $\psi_m=80^\circ$, and \edt{(ii)} a pitching motion \edt{defined as} $\theta(t)=\theta_m\tanh[C_{\eta}\sin(2\pi ft)]/\tanh(C_{\eta})$, where $\theta_m=45^\circ$ and $C_{\eta}=3.3$. The pitching axis is located at \edt{a distance of} $0.16c$ from the leading edge, while the stroke rotation is defined about a vertical axis positioned \edt{at a gap of} $0.4c$ from the wing\edt{'s} root. The Reynolds number is fixed at $Re=100$, based on the average \edt{velocity of the tip of the wing}, \edt{which follows } the benchmark configuration. \edt{It is important to mention that we keep} the computational domain, geometry \edt{of the wing}, and \edt{the} kinematics identical to those reported by Suzuki et al.~\cite{Suzuki_Minami_Inamuro_2015}. \edt{We perform the} simulation for four flapping cycles, and the instantaneous $C_L$ and $C_D$ obtained during the fourth cycle are used for validation. Figure~\ref{fig:Validation} presents a comparison of the predicted aerodynamic force coefficients with \edt{the results previously reported in different research investigations \cite{Suzuki_Minami_Inamuro_2015, medina2015illustration, dilek2019numerical, lionetti2022aerodynamic}}. \edt{The comparison of plots presented} in Figs.~\ref{fig:Validation_CL} and~\ref{fig:Validation_CD} clearly \edt{exhibits that our computational methodology} accurately \edt{reproduces} the temporal \edt{profiles of} both $C_L$ and $C_D$ throughout the flapping cycle. The predicted force histories capture the characteristic peaks, troughs, and phase variations with good agreement with previously published results \edt{in Refs.} \cite{Suzuki_Minami_Inamuro_2015, medina2015illustration, dilek2019numerical, lionetti2022aerodynamic}. \edt{This solver was also extensively used for research investigations related to a broad set of fluid-structure interactive systems \cite{farooq2025accurate, kamran2025role, fardi2025characterizing, maleksabet2026vortex, fardi2026fully}.} 

\section{Results and Discussion}
\label{sec:result}
The primary objective of this \edt{research} is to examine the influence of three-dimensional vortex dynamics on the aerodynamic performance of three different \edt{geometries of avian wings}. Particular emphasis is placed on identifying the relationship between the vortex structures \edt{produced by the flapping wings} and the resulting aerodynamic forces. To \edt{attain} this objective, computational simulations are performed for three configurations, denoted as $\mbox{Bird~1}$, $\mbox{Bird~2}$, and $\mbox{Bird~3}$. The simulations cover a range of governing kinematic and flow parameters, as summarized in Table~\ref{tab:parameters}. \edt{In this work,} {Strouhal number is defined as}~\textcolor{black}{\cite{joshi2020variational}}: 

\begin{equation}
\mbox{St}=\frac{2{A}{f}}{U_{\infty}}
\end{equation}

\noindent where $A$ is the maximum one-sided amplitude \edt{for a flapping wing}, and $f$ \edt{denotes} the flapping frequency. A representative free-stream velocity of $U_{\infty}=12.5~\mathrm{m\,s^{-1}}$ ($45~\mathrm{km\,h^{-1}}$) is considered \edt{relevant to real} flight \edt{conditions}\textcolor{black}{~\cite{CornellLab}}. \edt{As we solve the nondimensional form of the governing equations for fluid dynamics, we keep} $U_{\infty}=1~\mathrm{m\,s^{-1}}$ \edt{in our simulations, whereas} $A$ remains constant. To preserve the corresponding Strouhal numbers, the flapping frequency is scaled according \edt{to the relevant $\mbox{St}$}. Accordingly, the physical flapping frequencies of $4~\mathrm{Hz}$, $5~\mathrm{Hz}$, and $6~\mathrm{Hz}$ are scaled to $0.32~\mathrm{Hz}$, $0.40~\mathrm{Hz}$, and $0.48~\mathrm{Hz}$, respectively. These conditions correspond to the investigated \edt{conditions related to} $\mbox{St}=0.18$, $0.225$, and $0.27$ \edt{here}. 

\begin{table}[H]
\centering
\caption{\edt{Values of parameters used in our current work}}
\label{tab:parameters}
\begin{tabular}{lc}
\hline
\textbf{Parameter} & \textbf{Value} \\
\hline
Strouhal number ($\mbox{St}$) & $0.18$, $0.225$, $0.27$ \\
Reynolds number ($\mbox{Re}$) & $2500$, $5\times{10^3}$, $10^4$ \\
Bird configurations & $\mbox{Bird~1}$, $\mbox{Bird~2}$, $\mbox{Bird~3}$ \\
Wing flapping amplitude ($A$) & $0.5635$ \\
Flapping frequency ($f$) & $4\text{ - }6~ \mbox{Hz}$ \\
Free-stream velocity ($U_{\infty}$) & $12.5~\mbox{m/s}$ \\
Normalized wingspan ($S$) & $1$ \\
\hline
\end{tabular}
\end{table}

\edt{We begin our analysis by examining $\bar{C_D}$ shown in Figure~\ref{fig:mean_CD}} to compare the aerodynamic performance of the three configurations of wing-body combinations \edt{for different $\mbox{Re}$ and $\mbox{St}$}. \edt{Under all conditions}, $\mbox{Bird}~2$, \edt{having wings with serrations only,} \edt{experiences} the highest $\bar{C_D}$ among the three configurations. \edt{It is also important to observe that $\mbox{Bird}~1$, having flat wings, have the least $\bar{C_D}$, except when $\mbox{Re}=10^4$ and $\mbox{Re}=5\times10^3$, $\mbox{St}=0.18$, where $\mbox{Bird}~3$ has the least $\bar{C_D}$.} \edt{These observations appear consistent with the findings recently reported by Alenius et al. \cite{alenius2026feather}, where they concluded that} feather-like structures could produce aerodynamic lift comparable to conventional \edt{engineered airfoils}, \edt{but with some aerodynamic drag penalty}. This observation indicates that the geometric characteristics associated with \edt{serrated} feather\edt{s-like} morphology \edt{at the trailing part of the wing} \edt{could significantly} influence the aerodynamic drag by up to \edt{$9.19\%$}, \textcolor{black}{seen in our current work. The corresponding percentage differences in $\overline{C_D}$, calculated with respect to $\mbox{Bird~1}$ at each Reynolds number and Strouhal number, are summarized in Table~\ref{tab:CD_percentage}. $\mbox{Bird~2}$ exhibits a higher drag than $\mbox{Bird~1}$ by approximately $3.98\%$--$9.19\%$, with the maximum difference of $9.19\%$ occurring at $\mbox{St}=0.225$ and $\mbox{Re}=5\times{10^3}$. In comparison, the percentage difference between Bird~3 and Bird~1 remains smaller, ranging from approximately $0.58\%$ to $5.04\%$. Thus, although the incorporation of the feather morphology in $\mbox{Bird~2}$ increases the drag relative to $\mbox{Bird~1}$, the additional feather arrangement in $\mbox{Bird~3}$ generally reduces the drag penalty observed for $\mbox{Bird~2}$. Furthermore, the percentage difference does not exhibit a monotonic dependence on $\mbox{Re}$, indicating that the relative difference in drag varies with both the Reynolds number and flapping frequency.}

\begin{table}[ht]
\centering
\caption{\edt{Percentage difference in $\bar{C_D}$ relative to $\mbox{Bird}~1$ at different Reynolds numbers and Strouhal number}}
\label{tab:CD_percentage}
\renewcommand{\arraystretch}{1.2}
\begin{tabular}{ccccc}
\hline
\textbf{$\mbox{St}$} 
& \textbf{Configuration}
& $\boldsymbol{\mbox{Re}=2500}$
& $\boldsymbol{\mbox{Re}=5\times{10^3}}$
& $\boldsymbol{\mbox{Re}=10^4}$ \\
\hline

\multirow{3}{*}{$0.18~\mathrm{}$}
& Bird~1 & Reference & Reference & Reference \\
& Bird~2 & 8.03\% & 5.72\% & 3.98\% \\
& Bird~3 & 2.63\% & 1.34\% & 0.58\% \\
\hline

\multirow{3}{*}{$0.225~\mathrm{}$}
& Bird~1 & --- & --- & --- \\
& Bird~2 & 8.83\% & 9.19\% & 5.18\% \\
& Bird~3 & 3.29\% & 2.87\% & 0.76\% \\
\hline

\multirow{3}{*}{$0.27~\mathrm{}$}
& Bird~1 & --- & --- & --- \\
& Bird~2 & 7.59\% & 7.77\% & 7.77\% \\
& Bird~3 & 2.89\% & 2.55\% & 5.04\% \\
\hline

\end{tabular}
\end{table}

 \edt{Nevertheless, natural avian feathers with organized \edt{secondary} structural features, including herringbone arrangements of feather barbs, can {potentially reduce} aerodynamic drag for birds \cite{chen2013biomimetic, aditya2019recent}. Although these \edt{secondary and tertiary layers of individual feathers} are not explicitly resolved in \edt{our} presently reconstructed geometries}, the differences in \edt{$\bar{C_D}$ for} $\mbox{Bird}~2$ and $\mbox{Bird}~3$ indicate that the physiological arrangement of the feathered regions \edt{over the whole span of the wings substantially} influence their aerodynamic drag. 
 

\begin{figure}[H]
    \centering
    \includegraphics[width=1.0\textwidth]{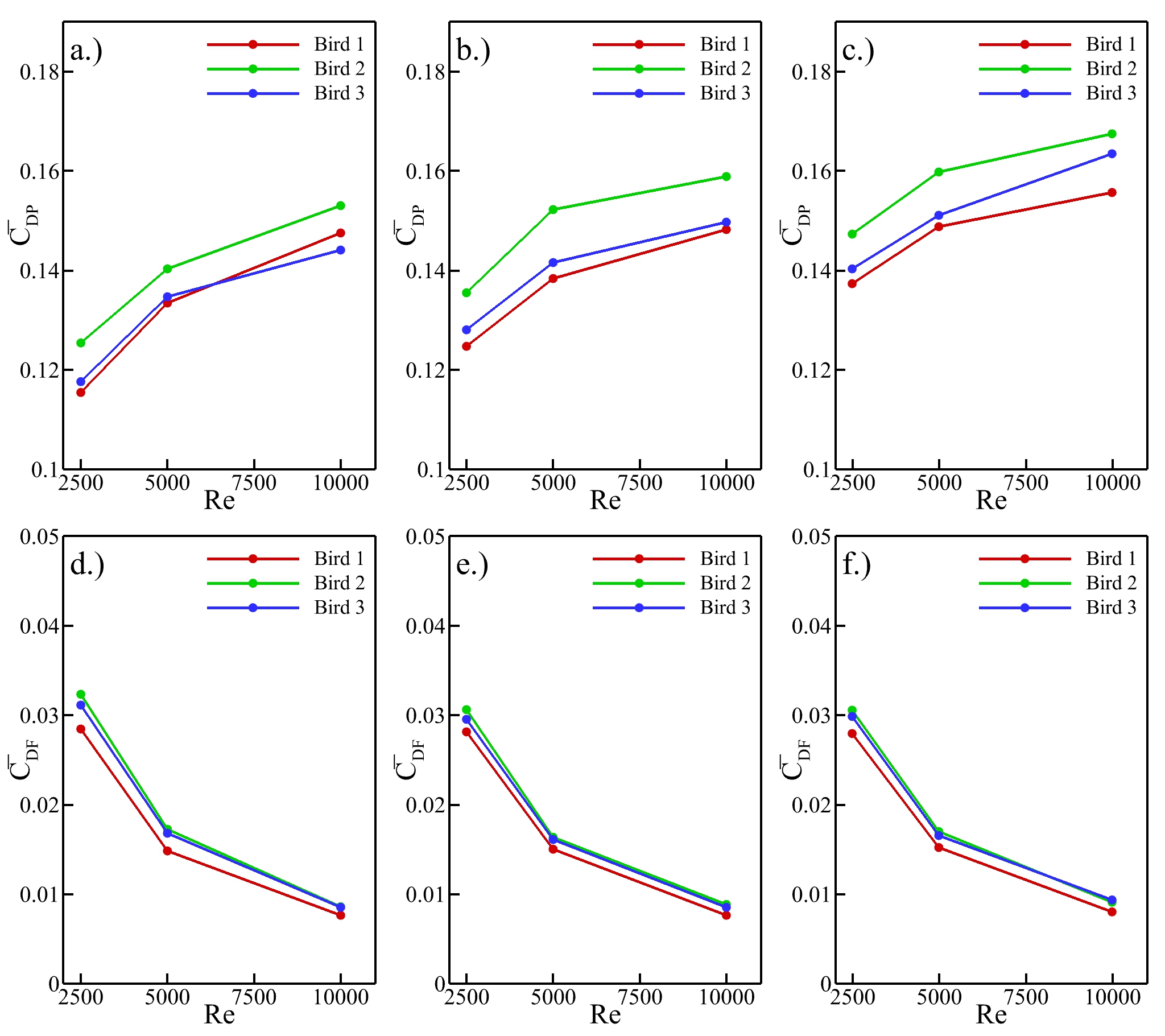}
    \caption{Variation in the mean pressure-drag coefficient ($\bar{C_{D_P}}$), and mean viscous-drag coefficient ($\overline{C_{D_F}}$), with $\mbox{Re}$ for the three bird configurations at different $\mbox{St}$. Plots (a) -- (c) show 
    $\overline{C_{D_P}}$ at (a) $\mbox{St}=0.18$, (b) $\mbox{St}=0.225$, and (c) $\mbox{St}=0.27$, respectively. Plots (d) -- (f) present $\overline{C_{D_F}}$ at (d) $\mbox{St}=0.18$, (e) $\mbox{St}=0.225$, and (f) $\mbox{St}=0.27$, respectively.}
    \label{fig:pressure_viscous_drag}
\end{figure}

\textcolor{black}{To further examine the contributions to the mean drag, 
Fig.~\ref{fig:pressure_viscous_drag} presents the pressure and viscous components of $\overline{C_D}$ for the three bird configurations. As shown in Figs.~\ref{fig:pressure_viscous_drag}a, \ref{fig:pressure_viscous_drag}b, and \ref{fig:pressure_viscous_drag}c, the mean pressure-drag coefficient, $\overline{C_{D_P}}$, increases with an increasing $\mbox{Re}$ for all three configurations and $\mbox{St}$. Among the three configurations, $\mbox{Bird}~2$ consistently exhibits the highest $\overline{C_{D_P}}$, whereas the difference between $\mbox{Bird}~1$ and $\mbox{Bird}~3$ remains comparatively smaller. The increase in $\overline{C_{D_P}}$ with $\mbox{St}$ is also evident when comparing the plots in Fig.~\ref{fig:pressure_viscous_drag}. On the contrary, Figs.~\ref{fig:pressure_viscous_drag}d, \ref{fig:pressure_viscous_drag}e, and \ref{fig:pressure_viscous_drag}f show that the mean viscous-drag coefficient, $\overline{C_{D_F}}$, decreases substantially with \edt{an} increasing $\mbox{Re}$ for all three configurations. The largest reduction occurs between $\mbox{Re}=2500$ and $\mbox{Re}=5\times{10^3}$, followed by a further decrease at $\mbox{Re}=10^4$. The differences in $\overline{C_{D_F}}$ among the three bird configurations remain relatively small compared with those observed for $\overline{C_{D_P}}$. Furthermore, the magnitude of $\overline{C_{D_P}}$ is considerably larger than that of $\overline{C_{D_F}}$ over the investigated conditions. These results indicate that the differences in the overall mean drag among the three bird configurations are predominantly associated with differences in the pressure-drag contribution, particularly for $\mbox{Bird}~2$.}







\begin{figure}[H]
    \centering
    \includegraphics[width=1.0\textwidth]{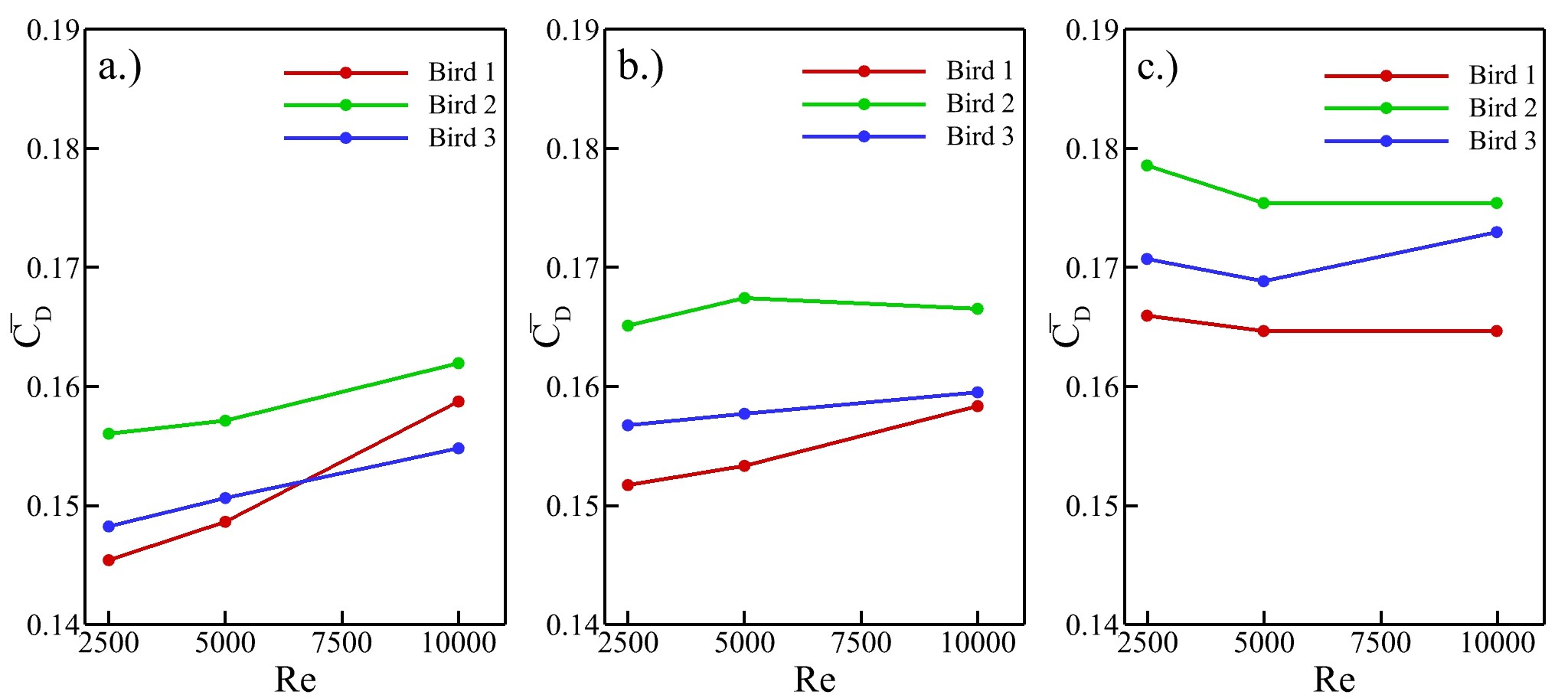}
    \caption{{Variations} \edt{in $\bar{C_D}$} with \edt{$\mbox{Re}$}
    at $\mbox{St}$ of (a) \edt{$0.18$, (b) $0.225$, and (c) $0.27$}}
    \label{fig:mean_CD}
\end{figure}

\begin{figure}[H]
    \centering
    \includegraphics[width=1.0\textwidth]{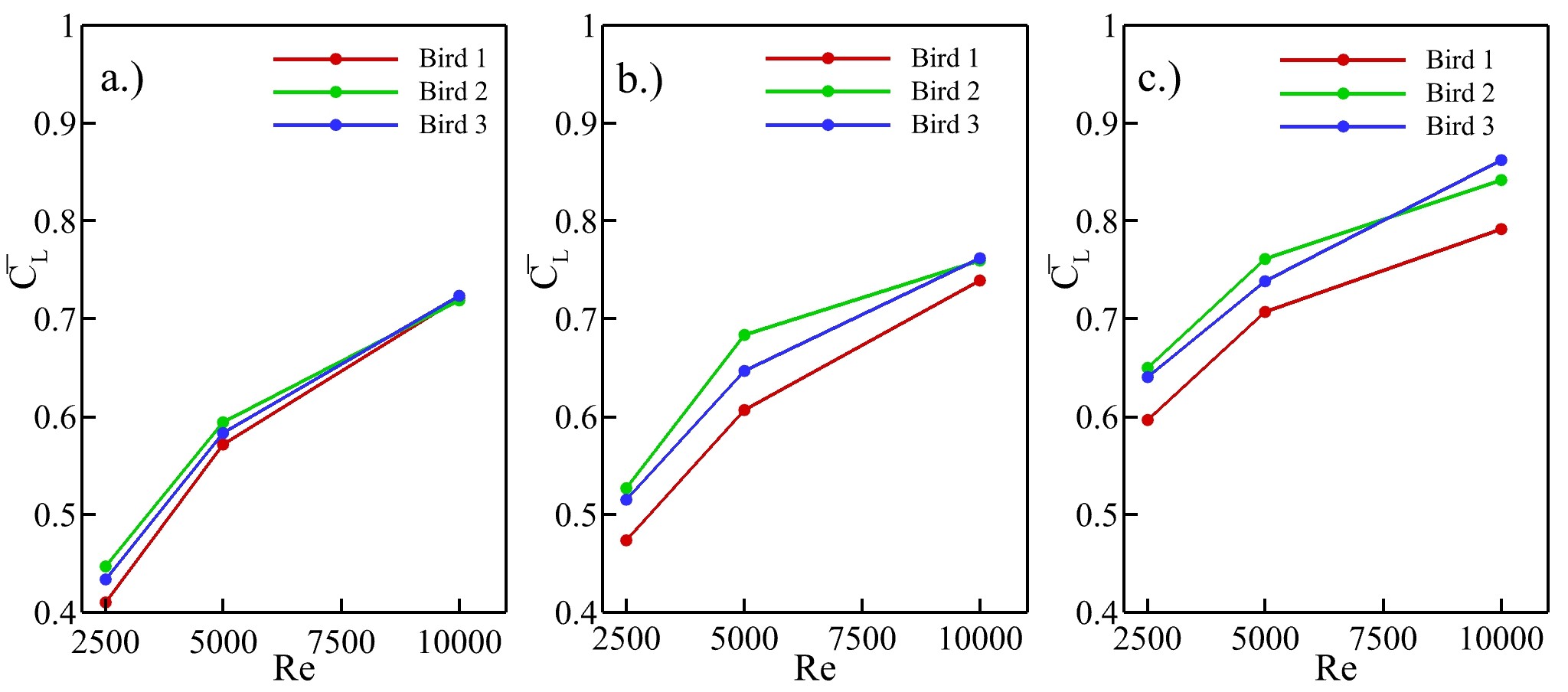}
    \caption{Variation in $\bar{C_L}$ with $\mbox{Re}$ at $\mbox{St}$ of (a) $0.18$, (b) $0.225$, and (c) $0.27$}
    \label{fig:mean_CL}
\end{figure}

\edt{Now, we analyze $\bar{C_L}$ for the three configurations of the avian wings using the plots presented in Fig.~\ref{fig:mean_CL}. It is observed that, in general, $\mbox{Bird}~1$ produces the minimum lift under all values of $\mbox{Re}$ and $\mbox{St}$. On the other hand, $\mbox{Bird}~2$, among all configurations, generates the maximum lift, but $\mbox{Bird}~3$ outperforms it for all $\mbox{St}$ at $\mbox{Re}=10^4$. It is also important to note that the difference in $C_L$ for $\mbox{Bird~2}$ and $\mbox{Bird~3}$ is only significant at $\mbox{Re}=5\times{10^3}$. This time-averaged data demonstrates that having serrations and multi-layered feather-like geometric features over the span of the wings is beneficial for maximizing mean lift force.}




\vspace{\baselineskip}

\begin{figure}[H] 
\centering 
\includegraphics[width=1.0\textwidth]{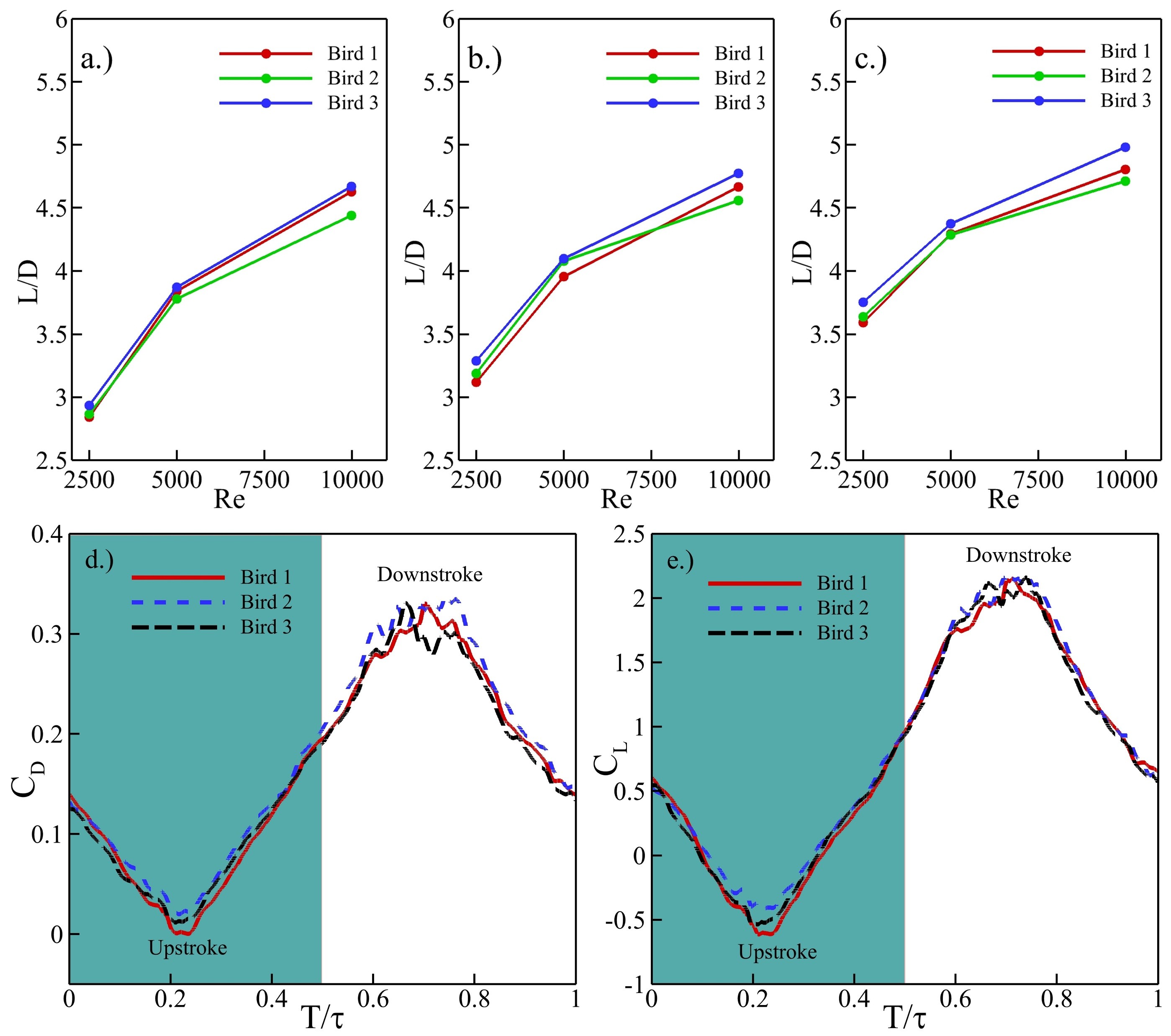} 
\caption{\edt{Comparison of} $L/D$ for the three bird configurations at different  $\mbox{Re}$ and $\mbox{St}$: (a) $\mbox{St}=0.18$, (b) $\mbox{St}=0.225$, and (c) $\mbox{St}=0.27$. Plots (d) \& (e) show variations of $C_L$ and $C_D$, respectively, over one oscillation cycle at the highest $\mbox{Re}$ and $\mbox{St}$} 
\label{fig:LD_CL_CD} 
\end{figure}

\edt{While the preceding observations somewhat shows conflicting roles of geometric variations in the avian wings for production of streamwise and lift forces, it is imperative to use the lift-to-drag ratio (${L}/{D}$) to make better conclusive assessments. We present this data in Fig.~\ref{fig:LD_CL_CD}, where ${L}/{D}$ is plotted versus $\mbox{Re}$ for three different values of $\mbox{St}$. Overall, $\mbox{Bird}~3$ demonstrates the superior aerodynamic performance in comparison to the other two configurations under all flow and kinematic conditions considered in this work. It is interesting to notice that $\mbox{Bird}~1$ and $\mbox{Bird}~2$ shows poor aerodynamic performance, mostly on the same levels, in terms of the lift-to-drag ratio. These findings motivates to investigate how the temporal profiles of these aerodynamic forces look. Hence, we choose to examine these instantaneous histories for all the three conditions at $\mbox{Re}={10^4}$ and $\mbox{St}=0.27$, as indicated by a small rectangle with dashed lines around the relevant data points in Fig.~\ref{fig:LD_CL_CD}c. So, Figs.~\ref{fig:LD_CL_CD}d and \ref{fig:LD_CL_CD}e
shows plots for $C_D$ and $C_L$, respectively. Although the profiles here apparently look alike for the three birds, yet there exist some subtle differences. For instance, the amplitudes for both aerodynamic force coefficients are greater for $\mbox{Bird}~1$ during the upstroke. In the downstroke that is primary lift-production motion for birds, $\mbox{Bird~3}$ attains the maximum amplitude for $C_D$ earlier than the other two configurations. More importantly, the peak region of $C_L$ for $\mbox{Bird 3}$ is wider, hinting at maintaining the maximum lift for a larger period of time. Hence, these observations provide us the basis for performing a deeper analysis on what happens in terms of vortex-vortex and vortex-wing interactions that could help birds improve their aerodynamic performance with variations in morphologies of their wings. }

\begin{figure}[H]
    \centering
    \includegraphics[width=\textwidth]{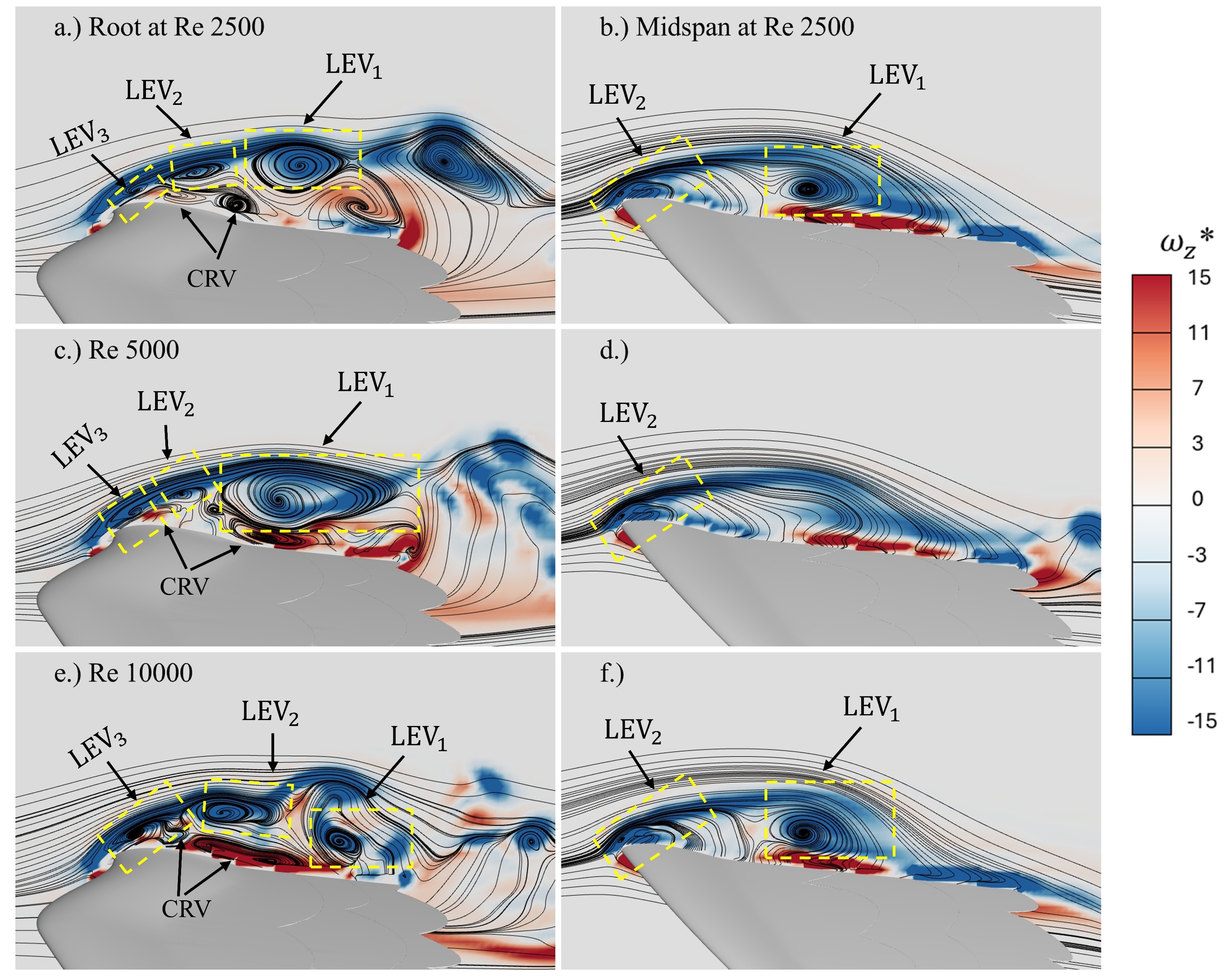}
    \caption{{Instantaneous LEV structures at the root} and midspan sections  of $\mbox{Bird}~3$ at $\mbox{St}=0.27$ during the downstroke: (a,b) $\mbox{Re}=2500$, (c,d) \edt{$\mbox{Re}=5\times{10^3}$}, and (e,f) \edt{$\mbox{Re}=10^4$}}
    \label{fig:LEV_PRESENCE}
\end{figure}

\edt{It is well known that leading-edge vortices play a dominant role in aerodynamics of insects and birds \cite{ellington1999novel, garmann2013three, jardin2014spanwise, muijres2008leading, hubel2010importance}. A few research studies also showed evidence for} formation of a dual-$\mbox{LEV}$ structure\edt{,} consisting of a stronger primary $\mbox{LEV}$ located farther inboard and a smaller minor $\mbox{LEV}$ that \edt{remained} closer to the leading edge. \edt{These flow structures exist over the wings} together with a counter-rotating vortex ($\mbox{CRV}$) associated with secondary flow separation beneath or in between the co-rotating $\mbox{LEV}$ \cite{muir2017leading,lu2006dual}. \edt{Additionally,} Muir et al.~\cite{muir2017leading} also \edt{reported} a triple-$\mbox{LEV}$ structure \edt{formed over a delta wing, like those of a common swift (\textit{Apus apus})} under certain flow conditions, where additional co-rotating vortical structures \edt{were formed} near the leading edge. \edt{At the instant where $\mbox{Bird}~3$ generates the maximum lift at $\mbox{St}=0.27$, Fig.~\ref{fig:LEV_PRESENCE} shows the existence of $\mbox{LEVs}$ through contour plots of spanwise vorticity ($\omega_z$) and streamlines of the flow at different sections of the wing. Figures~\ref{fig:LEV_PRESENCE}a, \ref{fig:LEV_PRESENCE}c, and \ref{fig:LEV_PRESENCE}e depict vortex dynamics on a plane near the root of the wing at $\mbox{Re}=2.5\times{10^3}$, $5\times{10^3}$, and $10^4$, respectively, and Fig.s~\ref{fig:LEV_PRESENCE}b, \ref{fig:LEV_PRESENCE}d, and \ref{fig:LEV_PRESENCE}f illustrate the flow features on the mid-plane over the wing for these conditions. We clearly observe that three $\mbox{LEVs}$, termed as $\mbox{LEV}_1$ (the largest one on the right), $\mbox{LEV}_2$ (the mid-level structure), and $\mbox{LEV}_3$ (the smallest one on the leading-edge of the wing) are developed. These labels distinguish the simultaneously occurring structures and facilitate tracking of these vortices in our subsequent analysis. It is pertinent to state that $\mbox{LEV}_1$ is detached from the surface of the wing at $\mbox{Re}=2.5\times{10^3}$ at this instant, but it is found closer to the wing at the higher values of $\mbox{Re}$. This characteristic may provide plausible reason behind greater lift with an increase in $\mbox{Re}$ due to the higher pressure difference due to this coherent flow structure. Coming to the middle of the span, $\mbox{LEV}_3$ does not develop itself to that location, and we only find $\mbox{LEV}_1$ and $\mbox{LEV}_2$ present there. Thus, a triple-LEV arrangement occurs near the root, while a dual-LEV arrangement persists at the midspan of the wing, which is consistent with the multiple-LEV topology reported by Muir et al. \cite{muir2017leading} and Lu et al. \cite{lu2006dual}. It is important to mention that $\mbox{Bird}~1$ and $\mbox{Bird}~2$ also show the existence of similar flow structure, which are not shown here for brevity.}

\edt{To further elucidate these findings,} $\mbox{LEV}_1$ is a well-developed coherent structure over the wing \edt{at $\mbox{Re}=2.5\times{10^3}$ in Fig.~\ref{fig:LEV_PRESENCE}a}, while the smaller $\mbox{LEV}_2$ and $\mbox{LEV}_3$ remain closer to the leading edge with the presence of two $\mbox{CRVs}$. An additional vortical structure occurs downstream of $\mbox{LEV}_1$ and farther from the wing surface, indicating a previously shed vortex. At the midspan \edt{in} Fig.~\ref{fig:LEV_PRESENCE}b, \edt{we notice} coexistence of $\mbox{LEV}_1$ and a weaker $\mbox{LEV}_2$, with a distinct gap between the two structures, which is less than what we observe at the root. A counter-rotating vortex is not clearly identified within this gap here. Its absence may contribute to the persistence of the co-rotating $\mbox{LEVs}$, since an opposing vortical structure between them can modify their interaction and influence their attachment \edt{with the surface. H}owever, this behavior also depends on relative strength\edt{, in terms of circulation} of the interacting vortices. At this location over the span, $\mbox{LEV}_1$ remains attached to the wing and \edt{the shear layer spanning from the leading edge of the wing continuously keep feeding it. Here, $\mbox{LEV}_2$ exists as} a comparatively weaker structure. 

For $\mbox{Re}=5\times10^3$, Fig.~\ref{fig:LEV_PRESENCE}c shows the presence of the LEVs, along with separated flow and a counter-rotating vortex however at $\mbox{Re}=5\times{10^3}$ in Fig.~\ref{fig:LEV_PRESENCE}d, the plot shows only a smaller developing vortical structure without a comparably large coherent \edt{$\mbox{LEV}_1$. Also}, a separated flow region with elevated vorticity is present near the wing surface\edt{. However}, this separated flow does not \edt{seems to} roll up into a distinct coherent vortex structure \edt{at this time instant}. Since the comparison represents a single time instant during the downstroke, the absence of a \edt{fully-formed} $\mbox{LEV}$ at the midspan does not exclude its formation a little later during the flapping cycle. This difference indicates a variation in the instantaneous LEV development between the root and midspan regions. Similar spanwise variations in development, attachment, and breakdown \edt{of $\mbox{LEVs}$} were also reported for flapping wings in \edt{Refs.} \cite{lu2006dual,lentink2009rotational}. At $\mbox{Re}=10^4$ in Fig.~\ref{fig:LEV_PRESENCE}\edt{e} shows several distinguishable vortical structures near the root, \edt{where $\mbox{LEV}_1$} is located farther downstream, while \edt{$\mbox{LEV}_2$ and $\mbox{LEV}_3$} are positioned progressively closer to the leading edge and exhibit comparatively larger sizes. \edt{These flow structure might owe their new arrangement to the greater inertial effects at this value of $\mbox{Re}$.} Multiple LEV structures are also observed at the midspan, as shown in Fig.~\ref{fig:LEV_PRESENCE}\edt{f}. \edt{Previously, Birch et al. \cite{birch2004force} indicated such changes in the structure of a $\mbox{LEV}$} with increasing Reynolds number, where the negative-vorticity concentration rolls up more tightly and subsequently divides into two distinct regions, resulting in the formation of two concentrated vortices. An additional \edt{$\mbox{CRV}$ was also} reported \edt{to exists} between the leading edge and the primary $\mbox{LEV}$ \cite{birch2004force}. Overall, Fig.~\ref{fig:LEV_PRESENCE} confirms the presence of multiple LEVs at different values of $\mbox{Re}$. But, this analysis only establishes their occurrence at a \edt{specifically selected time} instant but does not characterize their complete temporal evolution. The following section examines the formation, growth, persistence, downstream convection, and shedding of the LEVs \edt{sequentially at different time instants during} the complete downstroke through a detailed analysis of their three-dimensional structures and evolution along span \edt{of the wing}.

Following the identification of multiple LEVs in the sectional analysis presented in Fig.~9, their three-dimensional evolution is examined throughout the complete downstroke cycle. \edt{Now, our} analysis focuses on \edt{comparing} all three \edt{bird-like} configurations \edt{by characterizing} the influence of \edt{their wings' geometries with and without serrations and feather-like structures} on the vortex dynamics at $\mbox{Re}=10^4$ and $\mbox{St}=0.27$. The development of $\mbox{LEV}_1$, $\mbox{LEV}_2$, and $\mbox{LEV}_3$ is tracked from the root toward the tip, with particular attention to their growth, coherence, attachment, interaction, spanwise development, and shedding. Their attachment to and movement away from the \edt{wings' surfaces}  are also examined in relation to the aerodynamic performance of the three configurations. \edt{From all the time period of the downstroke}, particular attention is given to the instant corresponding to the maximum lift coefficient \edt{(}$C_{L,\max}$\edt{)} to examine the spatial organization of the vortical structures associated with the maximum \edt{aerodynamic performance metrics}.


The three-dimensional coherent vortex structures are identified using the $Q$-criterion, defined as
\begin{equation}
Q=\frac{1}{2}\left(\|\boldsymbol{\Omega}\|^2-\|\mathbf{S}\|^2\right)>0,
\end{equation}

\noindent where strain-rate tensor \edt{($\mathbf{S}$)} and rotation-rate tensor \edt{($\boldsymbol{\Omega}$)} are expressed as:

\begin{equation}
\mathbf{S}
=\frac{1}{2}\left[\nabla\mathbf{u}+(\nabla\mathbf{u})^{T}\right],
\qquad
\boldsymbol{\Omega}
=\frac{1}{2}\left[\nabla\mathbf{u}-(\nabla\mathbf{u})^{T}\right],
\end{equation}

The $Q$-criterion distinguishes vortex-dominated regions from strain-dominated regions by comparing the local rotation rate with the strain rate. A positive value of $Q$ indicates that the local rotation dominates the strain and\edt{,} therefore\edt{,} identifies regions associated with vortical motion \edt{in a flow field}. Accordingly, connected regions of positive $Q$ represent the spatial extent and organization of coherent vortex structures. In \edt{our} present analysis, the three-dimensional coherent vortex structures are visualized using isosurfaces at a nondimensional threshold of $Q=10$.

\begin{figure}[H]
    \centering
    \makebox[\textwidth][c]{%
        \includegraphics[width=1.0\textwidth]{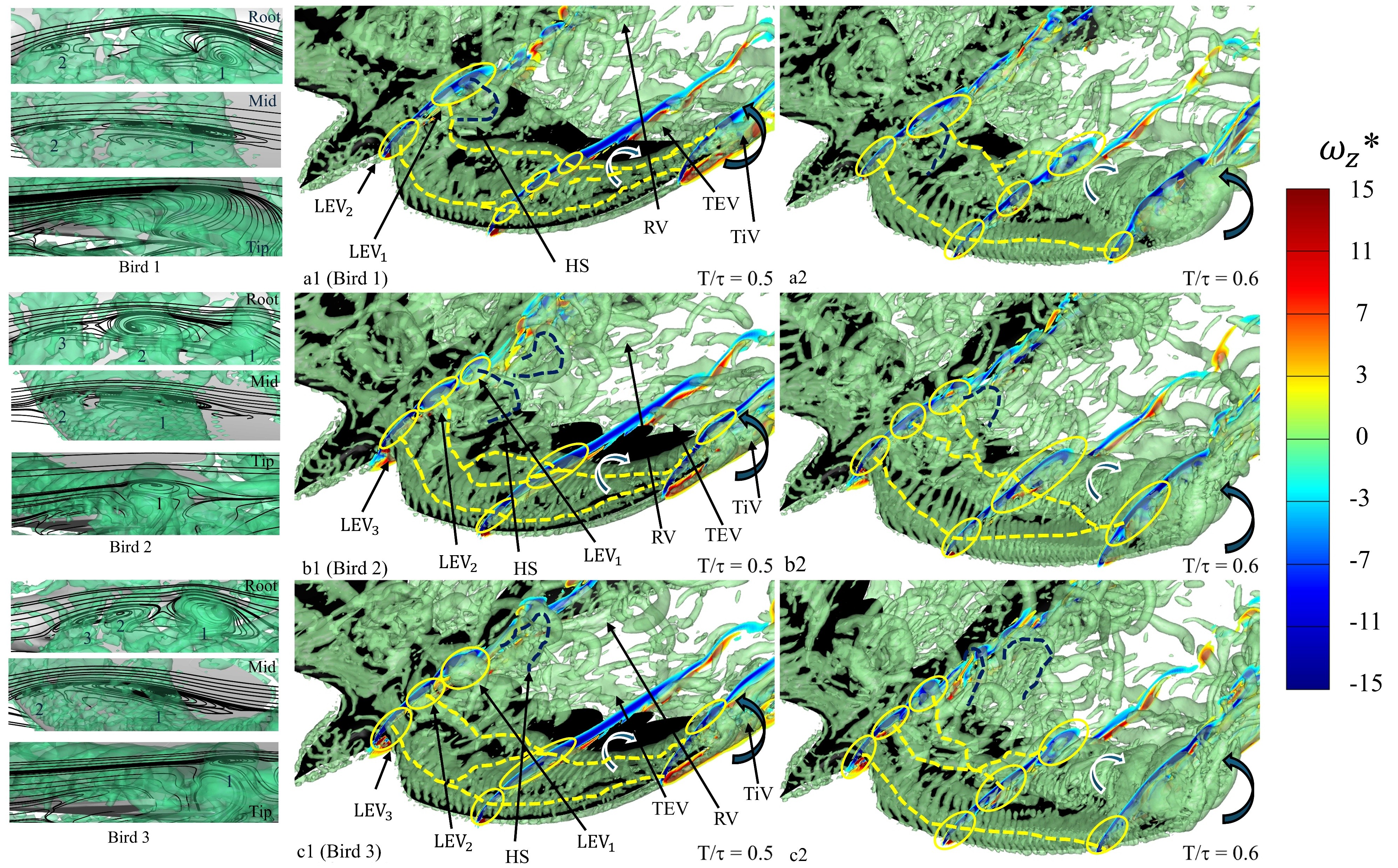}
    }
    \caption{\edt{Formation of $\mbox{LEVs}$} and \edt{their} spanwise connectivity for the three bird-\edt{like physiologies} at $T/\tau=0.5$ \edt{and} $0.6$ during the early downstroke at $\mbox{Re}=10^4$, visualized using $Q=10$}
    \label{fig:0-21}
\end{figure}


Figure~\ref{fig:0-21} combines the three-dimensional \edt{vortex} visualization with sectional vorticity contours and corresponding streamline fields to characterize the spanwise evolution of the LEVs. Cross-sectional planes are examined at selected locations near the root, midspan, and regions \edt{near the tip of the wings}. Yellow circles identify \edt{presence of} the \edt{vortices}, while the dashed lines indicate the spanwise connectivity and \edt{spanwise coherence} of the $\mbox{LEVs}$ from the root toward the wing\edt{'s} tip. These sectional flow fields enable the identification and tracking of individual vortical structures, primarily $\mbox{LEV}_1$, $\mbox{LEV}_2$, and $\mbox{LEV}_3$, and characterize their size, coherence, spatial position, attachment, and development along the span. The root vortex ($\mbox{RV}$), trailing-edge vortex ($\mbox{TEV}$), and tip vortex ($\mbox{TiV}$) are also identified\edt{. However}, this analysis primarily focuses on the $\mbox{LEVs}$ due to their dominant contribution to \edt{production of} lift under unsteady flapping conditions. \edt{The main aim here is to examine variations in this whole flow dynamics over the wing-body configurations potentially governed by different physiological features of the avian wings.}



\edt{At} $T/\tau=0.5$--$0.6$ in \edt{Fig.}~\ref{fig:0-21}1\edt{a}, \edt{where $\tau$ denotes the time period of a flapping cycle}, $\mbox{Bird}~1$ shows the formation of $\mbox{LEVs}$ near the root, where $\mbox{LEV}_1$ is located around the mid-chord region, while $\mbox{LEV}_2$ is located closer to the leading edge. In comparison, {Figs.~\ref{fig:0-21}b and \ref{fig:0-21}c} shows three vortical structures for $\mbox{Bird}~2$ and $\mbox{Bird}~3$, denoted as \edt{$\mbox{LEV}_1$}, \edt{$\mbox{LEV}_2$}, and \edt{$\mbox{LEV}_3$}, which are distributed over the chord. Near the root region of $\mbox{Bird}~1$, \edt{detachment of $\mbox{LEV}_1$ occurs earlier that further gets developed} into a horseshoe-like vortical structure, as indicated by the blue dashed line. A similar horseshoe-like vortex structure is also reported in a computational study \edt{related to a bat's} flight by Kumar et al. \cite{kumar2025computational}. Similar structures are also observed for $\mbox{Bird}~2$ and $\mbox{Bird}~3$\edt{. Nevertheless}, $\mbox{Bird}~1$ exhibits earlier \edt{detachment of the $\mbox{LEV}$} and greater flow unsteadiness, whereas $\mbox{Bird}~2$ and $\mbox{Bird}~3$ exhibit more developed horseshoe-like vortical structures forming in a series. \edt{Contrarily}, the smaller structures observed for $\mbox{Bird}~1$ show characteristics associated with the breakdown of the horseshoe-like vortical structure. \edt{P}ossible formation\edt{s} of these \edt{$\mbox{HVs}$} occur near the wing-body junction, where the flow over the curved body interacts with the flow over the wing. The curvature of the body redirects the local flow toward the wing, where it interacts with the $\mbox{LEV}$ extending in the spanwise direction toward the wing-tip region. This interaction produces a local disruption in the vortex structure, resulting in the formation of a curved, horseshoe-like vortex near the wing-body junction.

At the midspan in Fig.~\ref{fig:0-21}\edt{a1}, three smaller vortex structures are present for $\mbox{Bird}~1$, whereas comparatively larger vortex structures are observed for $\mbox{Bird}~2$ and $\mbox{Bird}~3$ shown in Fig.~\ref{fig:0-21}\edt{b1 and \ref{fig:0-21}c1}. These structures are also evident in the corresponding streamline fields presented on the left side of the figure. Near the wing-tip region in Fig.~\ref{fig:0-21}\edt{a1}, $\mbox{Bird}~1$ does not exhibit a clear roll-up of the shear layer into a coherent vortex structure, whereas a distinct roll-up is observed for $\mbox{Bird}~2$ and $\mbox{Bird}~3$ in \edt{Figs}.~\ref{fig:0-21}\edt{b1 and \ref{fig:0-21}c1} \edt{a little farther} from the {leading} edge, as slotted winglets are present. At this instant, the wing also forms a curved configuration toward the tip region. The separated primary feathers associated with the slotted winglets can spread the wing-tip vorticity among multiple vortex structures rather than concentrating it into a single tip-vortex core\edt{. Such observations are} consistent with the vortex-spreading mechanism for slotted \edt{wings} reported \edt{earlier} \cite{kleinheerenbrink2017multi}\edt{,} which could be the possible reason on forming those small \edt{structures,} as interaction of $\mbox{LEV}$ and $\mbox{TiV}$ \edt{could be} hindered during this \edt{time}. In addition, the dominant flow structures extend over more than half of the wing\edt{'s} span at this instant, with comparatively low \edt{disturbances in the flow}, and remain primarily {concentrated near the leading-edge region} \edt{for $\mbox{Bird}~2$ and $\mbox{Bird}~3$}. As the spanwise position progresses from the root toward the tip, the \edt{these coherent flow structures become narrower. However}, the flow remains relatively smooth and organized, with no pronounced increase in unsteadiness over the inner and midspan regions. Toward the outer wing, the coherent structures occupy a progressively smaller region for $\mbox{Bird}~2$ and $\mbox{Bird}~3$ \edt{(see Figs.}~\ref{fig:0-21}\edt{b1} and \ref{fig:0-21}\edt{c1} as they approach the wing\edt{'s} tip. At the same \edt{time}, the instantaneous $C_L$ continues to increase \textcolor{black}{(see Fig.~\ref{fig:LD_CL_CD}e)}. Hence, the increase in lift coincides with the development and persistence of coherent leading-edge vortex structures over the inner and midspan regions, while their spatial extent gradually decreases toward the wing tip.

At $T/\tau=0.6$, the flow structures over the wing cover a larger area compared with the previous time step, while coherent vortical structures near the root remain present for all three configurations \edt{in Figs.~\ref{fig:0-21}a2, \ref{fig:0-21}b2, and \ref{fig:0-21}c2}. \edt{On} the midspan, the size of the coherent structure increases slightly for $\mbox{Bird}~1$, whereas $\mbox{Bird}~2$ in Fig.~\ref{fig:0-21}\edt{b2} exhibits a larger vortex structure. The \edt{visible} increase in \edt{size of the vortex} could be associated with {viscous diffusion} \textcolor{black}{through which vorticity diffuses from the concentrated vortex core into the surrounding flow by transferring its momentum, resulting in a broader and more spatially distributed vorticity field.} \edt{H}owever, a {large-sized} $\mbox{LEV}$ does not necessarily represent an increase in circulation. For $\mbox{Bird}~3$ in Fig.~\ref{fig:0-21}\edt{c2}, \edt{$\mbox{LEV}_1$} splits into two distinct co-rotating vortex cores, accompanied by the presence of secondary separation near the wing\edt{'s} surface, which is also shown in the Fig.~\edt{\ref{fig:LEV_PRESENCE}}. Similar dual-vortex development \edt{was} associated with secondary separation in some previous \edt{research} studies\edt{, for example in Ref.~\cite{gordnier2003higher}}, where the interaction of the secondary separated flow with the primary shear layer leads to the formation of two same-\edt{signed} vortex structures. \edt{Moving} toward the region \edt{near the wing's tip} in \edt{Figs}.~\ref{fig:0-21}\edt{a2}, \ref{fig:0-21}\edt{b2}, and \ref{fig:0-21}\edt{c2}, the interaction between the $\mbox{LEVs}$, $\mbox{TiV}$ and $\mbox{TEV}$ forms a large swirl-like vortical structure prior to its breakdown\edt{,} and the flow inside its core becomes highly unsteady. At this \edt{time}, a smaller LEV is also observed near the tip region, which may be associated with the development of the larger interacting vortex structure. A noticeable difference is observed among the wing-tip configurations \edt{here}. For $\mbox{Bird}~1$, the big \edt{swirling} vortical structures near the tip merge into a comparatively large combined structure. \edt{But} for $\mbox{Bird}~2$ and $\mbox{Bird}~3$, where slotted winglets are present, the vortical structures remain distributed along the individual winglets and follow the slotted \edt{geometry of the wings at their tips} rather than forming a similarly concentrated combined structure \edt{that} looks like about to break or burst at \edt{during the} next \edt{half} stroke. 


\begin{figure}[H]
    \centering
    \makebox[\textwidth][c]{%
        \includegraphics[width=1.0\textwidth]{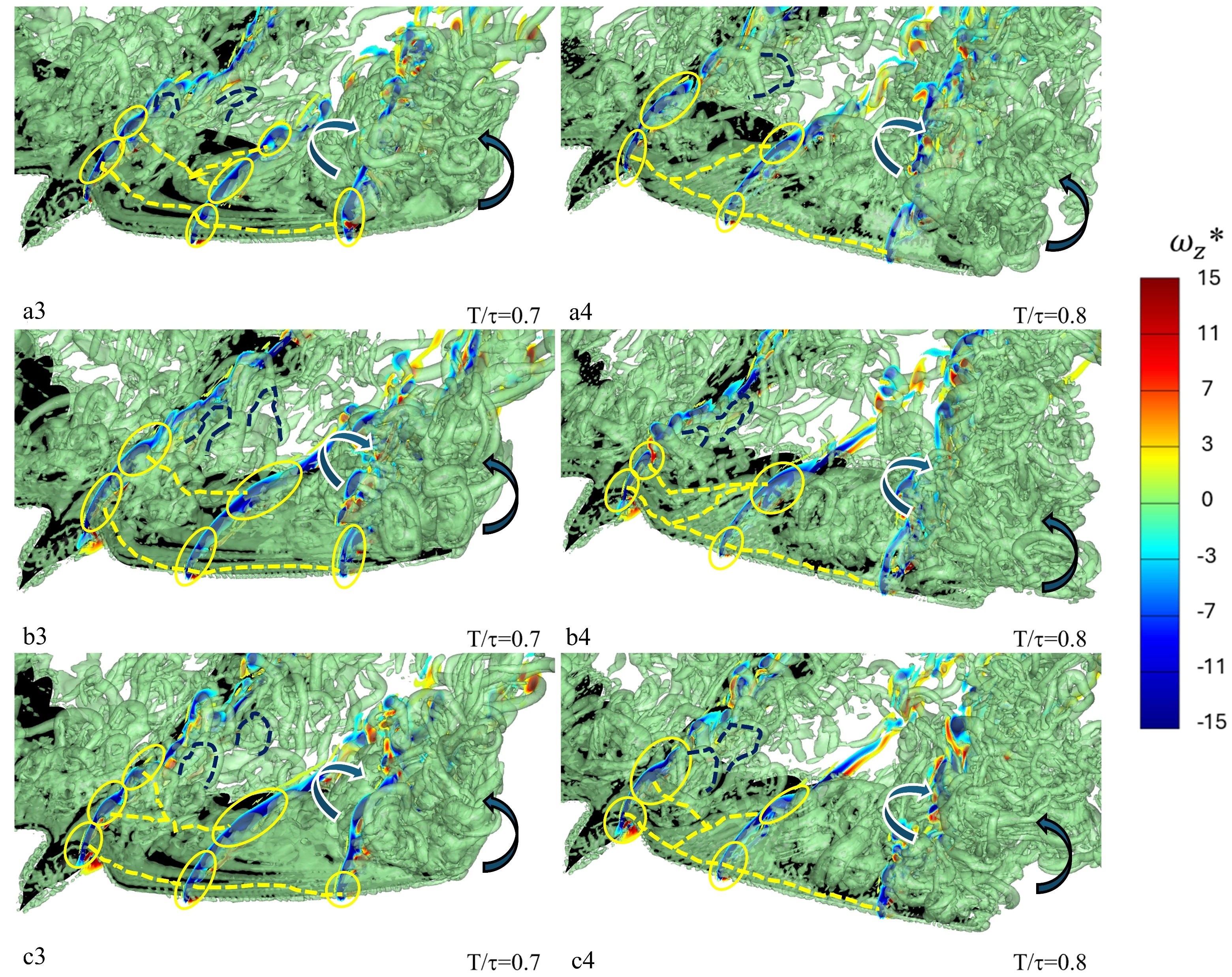}
    }
    \caption{\edt{Formation of $\mbox{LEVs}$} and \edt{their} spanwise connectivity for the three bird-\edt{like physiologies} at $T/\tau=0.7$ \edt{and} $0.8$ during downstroke at $\mbox{Re}=10^4$, visualized using $Q=10$}
    \label{fig:2-31}
\end{figure}

Figure~\ref{fig:2-31} presents the flow structures at $T/\tau=0.7$--$0.8$, corresponding to the stage of the mid-downstroke during which all three bird-wing configurations reach their maximum $C_L$. Similar LEV structures, \edt{as explained earlier here,} remain present near the root and midspan regions in \edt{Figs.}~\ref{fig:2-31}\edt{a3, \ref{fig:2-31}b3, and \ref{fig:2-31}c3}. However, the large swirl-like vortical structure observed at the preceding time instant shifts further toward the midspan and progressively loses its coherence near the wing-tip region. The effects of the local velocity \edt{of the wing} and \edt{its} acceleration also become more prominent \edt{at} this stage of the downstroke, contributing to the increased unsteadiness of \edt{these} flow \edt{features}. At $T/\tau=0.7$, a noticeable difference in \edt{development of} the flow is observed \edt{between} the three \edt{wing-body} configurations. Beyond the midspan region for $\mbox{Bird~1}$ in \edt{Fig.}~\ref{fig:2-31}\edt{a3}, the flow begins to lose its close contact with the wing\edt{'s} surface and extends toward the region where the vortical structures become increasingly disorganized, \edt{hinting at} the \edt{progressive breakdown of the vortex}. \edt{Oppositely}, the flow remains comparatively more \edt{attached} with the wing\edt{'s} surface from root to upto the midspan region at this instant for both $\mbox{Bird}~2$ and $\mbox{Bird}~3$ \edt{presented} in \edt{Figs.}~\ref{fig:2-31}\edt{b3 and \ref{fig:2-31}c3}. As the downstroke progresses toward $T/\tau=0.8$, the vortical region identified in the preceding time step increases in spatial extent and progressively loses its coherence and cover more area on the wing. The yellow dashed lines indicate the spanwise connection between the $\mbox{LEVs}$\edt{. H}owever, \edt{identifying} continuous distribution of coherent \edt{$\mbox{LEVs}$} become increasingly difficult at these time instants, \edt{because} \edt{interactions between} the {$\mbox{LEVs}$}, $\mbox{TEVs}$, and $\mbox{TiVs}$ intensify, resulting in the formation of smaller vortical structures and highly unsteady flow. This highly unsteady flow region \edt{traverses} from the wing\edt{'s} tip toward \edt{its} midspan in \edt{Figs.}~\ref{fig:2-31}\edt{a4, \ref{fig:2-31}b4, and \ref{fig:2-31}c4}. \edt{A careful look at} the three configurations \edt{tells about} a more rapid loss of coherent, surface-associated flow for $\mbox{Bird}~1$ and $\mbox{Bird}~2$ in \edt{Figs.}~\ref{fig:2-31}\edt{b3 and \ref{fig:2-31}c3} \edt{with more} unsteadiness toward the midspan. \edt{On the other hand,} $\mbox{Bird}~3$ retains more identifiable flow structures over a larger portion of the midspan {region of the wing, as exhibited} in \edt{Fig.}~\ref{fig:2-31}\edt{c3}. This difference is particularly evident from the flow fields over the midsection of the three wings at $T/\tau=0.8$.

\begin{figure}[H]
    \centering
    \makebox[\textwidth][c]{%
    \includegraphics[width=1.0\textwidth]{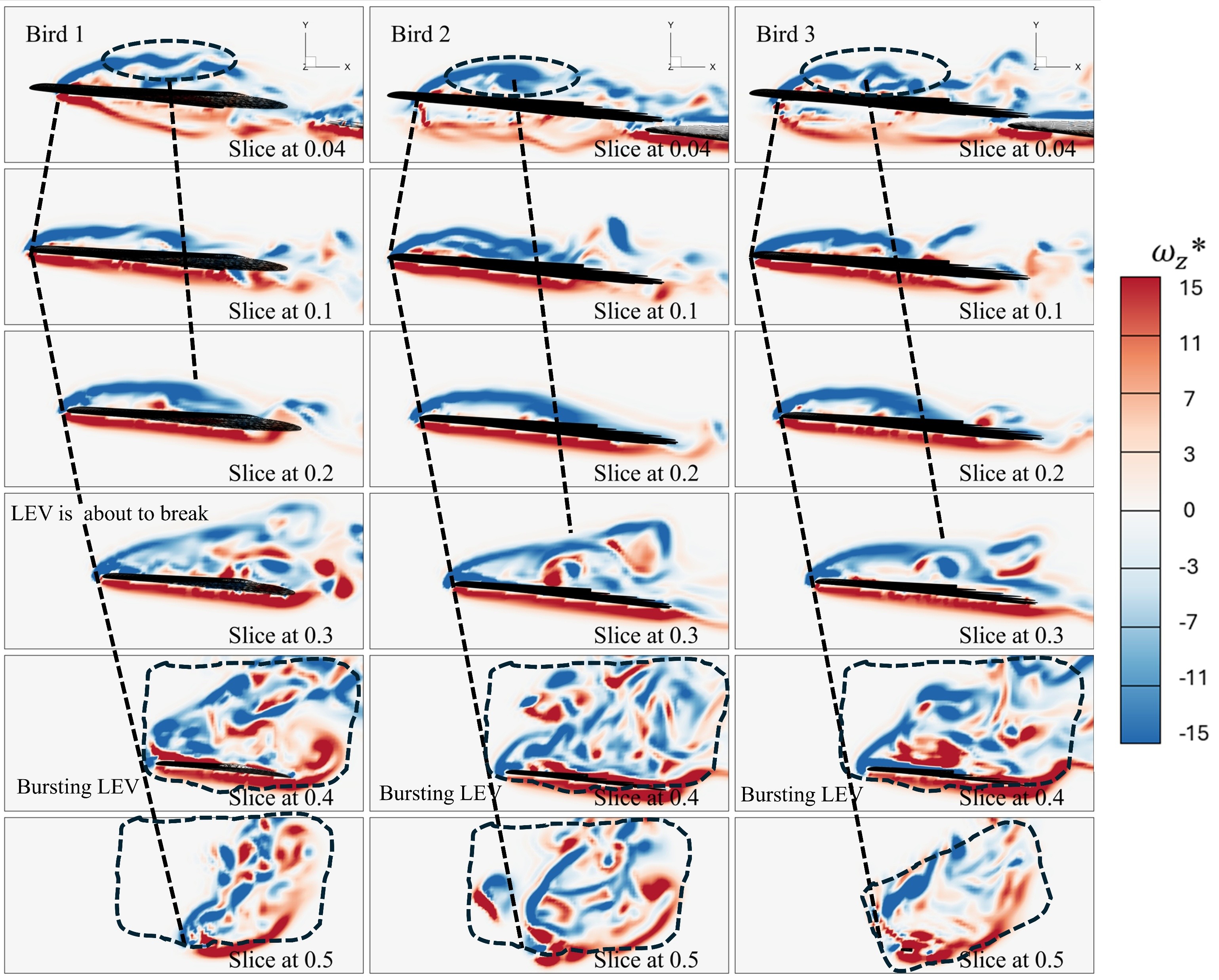}
    }
    \caption{\edt{Contours of $\omega_z$ over the different sections along the span of the three wings to compare the detachment process of flows over them at $T/\tau=0.7$}}
    \label{fig:LEV_Breakdown}
\end{figure}

\begin{figure}[H]
    \centering
    \makebox[\textwidth][c]{%
        \includegraphics[width=1.0\textwidth]{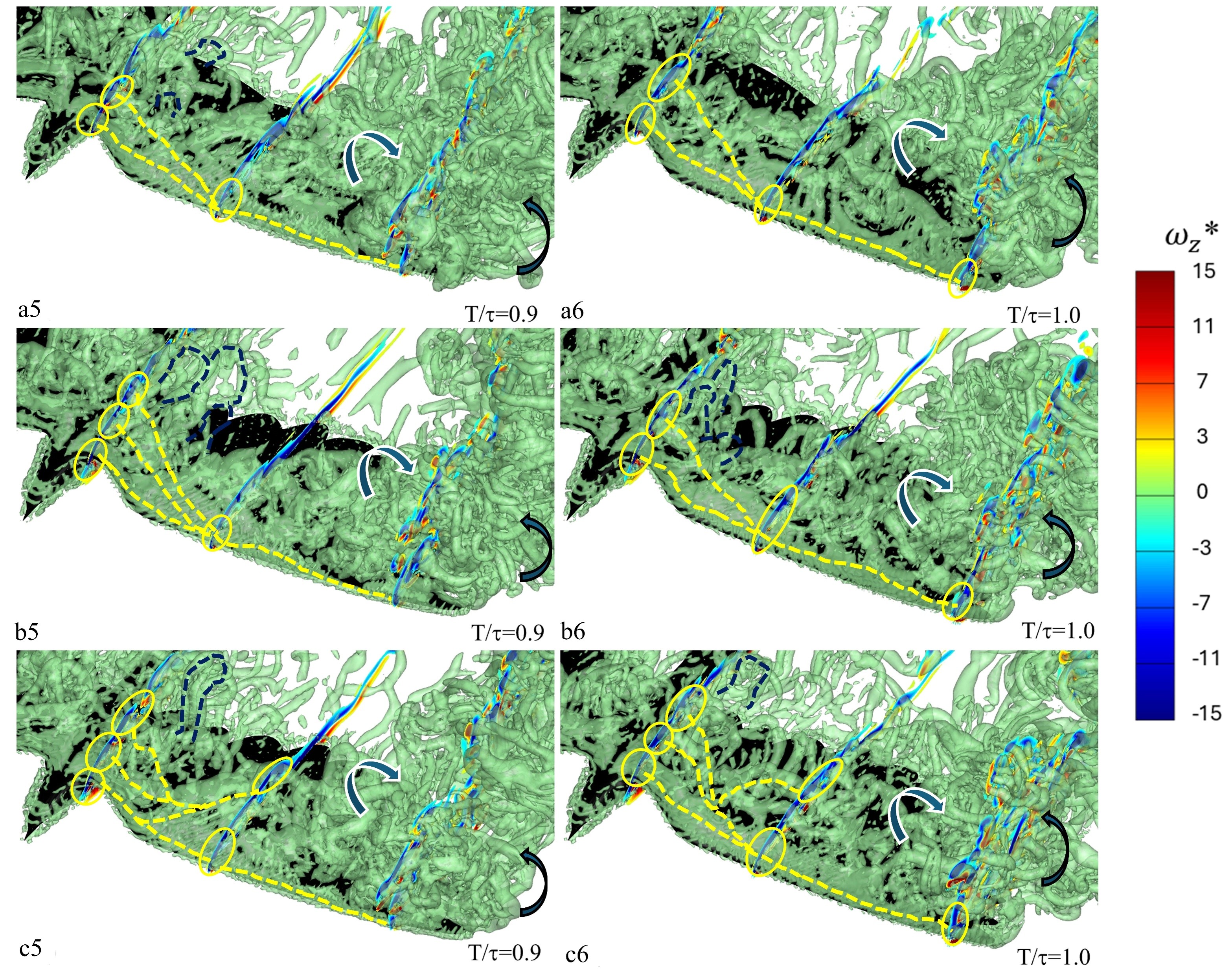}
    }
    \caption{\edt{Formation of $\mbox{LEVs}$} and \edt{their} spanwise connectivity for the three bird-\edt{like physiologies} at \edt{$T/\tau=0.9$ and $1.0$} during downstroke at $\mbox{Re}=10^4$, visualized using $Q=10$}
    \label{fig:5-61}
\end{figure}
 
To further examine this behavior, multiple sectional planes are created at $T/\tau=0.7$ shown in \edt{Fig.}~\ref{fig:LEV_Breakdown} at different spanwise locations to \edt{uncover behavior of} the \edt{flows} \edt{in terms of their attachment with the three wings' surfaces} and identify the spanwise \edt{locations} \edt{where} $\mbox{LEVs}$ breakdown. The first black dashed line, \edt{running down the columns of subplots here,} indicates the development of $\mbox{LEV}_2$ and/or $\mbox{LEV}_3$ at the corresponding \edt{time} instant, while the second black dashed line indicates \textcolor{black}{primary vortex $\mbox{LEV}_1$ and its} subsequent movement, coherence, and progression toward shedding or detachment from the wing\edt{'s} surface. The development and connectivity of these vortex structures are discussed above using the corresponding streamline patterns. The present sectional analysis primarily focuses on identifying the \edt{regions over the} wings\edt{' surfaces} over which these $\mbox{LEVs}$ remain present. At the spanwise slice location of $0.04$ \edt{({almost at the wing-body joint})}, \textcolor{black}{The primary vortex structure, $\mbox{LEV}_1$} in $\mbox{Bird}~1$ appears to move away from the wing surface, with a portion of the vortical structure already detached. \edt{Contrarily}, $\mbox{Bird}~2$ exhibits a larger coherent \textcolor{black}{$\mbox{LEV}_1$} over the wing, \edt{whereas} $\mbox{Bird}~3$ shows the presence of multiple LEVs that remain connected to the shear layer \edt{{rolling up at the leading edge}}. {At the spanwise location of} $0.1$, \edt{the} $\mbox{LEV}$ by $\mbox{Bird}~1$ \edt{appears to be recaptured by the wing before it gets} shed \edt{in the wake}, while the formation of a new LEV continues near the leading edge \edt{while coming closer to the wing at this stage}. \edt{This behavior is seen potentially due to the increased curvature of the wing while moving away from its root}. \edt{Comparing this process with with $\mbox{Bird}~1$}, $\mbox{Bird}~2$ and $\mbox{Bird}~3$ retain multiple LEV structures within \edt{the apparent continuous} shear \edt{layers} and in closer proximity to their \edt{relevant wings}. At the spanwise slice \edt{located} at $0.2$, all three configurations show the formation of large coherent \edt{flow} structure \edt{over the wings}. \edt{Interestingly, the $\mbox{LEV}$ for} $\mbox{Bird~1}$ \edt{undergoes its} breakdown \edt{at the} \edt{spanwise} location of $0.3$, where the initially coherent vortex structure \edt{gets burst} into several smaller structures, with a lower concentration of \edt{vorticity} near the leading edge. The flow also moves away from the wing\edt{'s} surface in this region, consistent with the behavior observed in the three-dimensional vortex dynamics presented in Fig~\ref{fig:2-31}\edt{a3}. \edt{Moving to the} spanwise locations of $0.4$ and $0.5$ {that represent} the near-tip \edt{regions}, the flow becomes increasingly unsteady. This behavior is associated with \edt{breakdown of $\mbox{LEVs}$} and \edt{their interactions with} $\mbox{TEVs}$, and $\mbox{TiVs}$. \edt{It is important to point out here that this unsteadiness} is more prominent \edt{in cases of} $\mbox{Bird~1}$ and $\mbox{Bird~2}$ than for $\mbox{Bird~3}$. \edt{Specifically}, $\mbox{Bird~3}$ \edt{experiences} comparatively maintains more coherent flow structures over \edt{its flapping wing} in the corresponding region. \edt{These subtle variations in the flow primarily owing to the physiological structures distributed over the span of the wing can be a significant contributing factor for increasing aerodynamic lift by $\mbox{Bird}~3$.}
 


\begin{figure}[H]
    \centering
    \makebox[\textwidth][c]{%
        \includegraphics[width=1.0\textwidth]{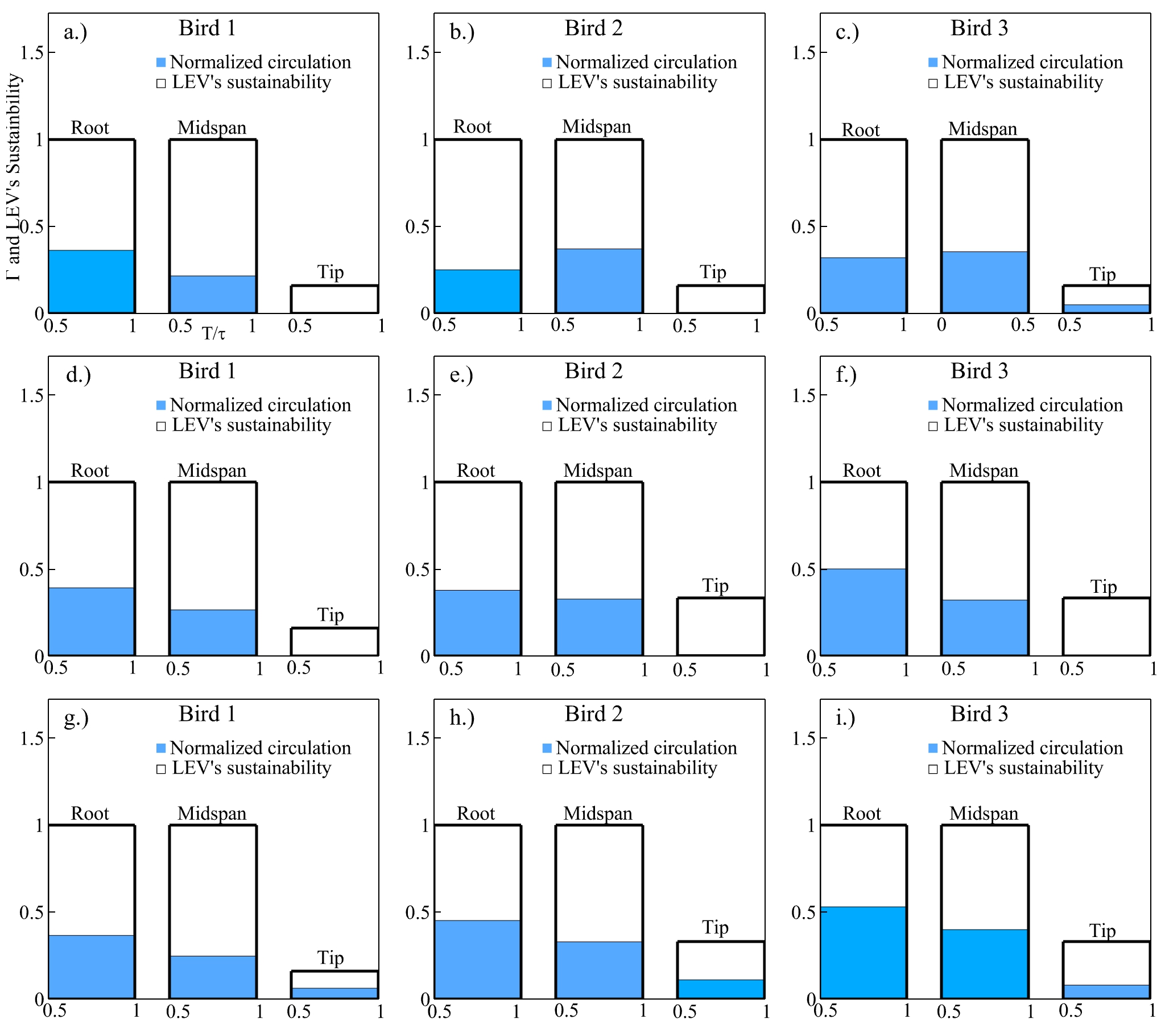}
    }
    \caption{Comparison of normalized circulation and sustainability \edt{of the $\mbox{LEVs}$} during the downstroke for the differnt wing configurations at $\mbox{Re}=10^4$ and  (a-c): $\mbox{St}=0.18$, (d-f): $0.225$, and (g-i): $0.27$}
    \label{fig:Root_mid_span_circ}
\end{figure}

\edt{Now, Fig.}~\ref{fig:5-61} presents the flow structures at $T/\tau=0.9$--$1.0$, \edt{corresponding to the time when downstroke is about to complete}. In \edt{Figs.}~\ref{fig:5-61}\edt{a5, \ref{fig:5-61}b5, and \ref{fig:5-61}c5}, all three configurations exhibit increased flow unsteadiness that extends beyond the midspan and progresses further toward the \edt{regions near the} root. Over the midspan region, distinct coherent vortex structures become increasingly difficult to identify, while the flow moves away from the wing\edt{'s} surface and extends further upward as the wing progresses downward, \edt{pointing to} progressive loss of \edt{coherence of LEVs}. \edt{One particular observation, evident from the plots at ${t/\tau}=1$ in Figs.~\ref{fig:5-61}a6, \ref{fig:5-61}b6, and \ref{fig:5-61}c6 relates to sustained coherence of flows near the root of the wing in each case, with $\mbox{Bird}~3$ still maintaining a \edt{clearly strong} LEV-like structure spanning over its whole leading edge to possibly create additional pressure difference for the wing to enhance its lift.}

\edt{Next, it is important to quantify strengths of the coherent flow structures over the three wing-body configurations of the bird-like physiologies under consideration in this work, so that their association with production of aerodynamic lift can be be further revealed. For this purpose, we compute circulations of the $\mbox{LEVs}$ along the span of the wings. Furthermore, we also evaluate their sustainability, in terms of the fractional time during the flapping period, when these structures remain clearly identifiable over different spanwise locations on the wings.} Figure~\ref{fig:Root_mid_span_circ} presents a quantitative comparison \edt{for these characteristics of} the $\mbox{LEVs}$\edt{, showing} the overall normalized circulation \edt{highlighted} with \edt{the} filled blue color and sustainability along the wing\edt{'s} span \edt{manifested by corresponding heights of} the rectangle boxes \edt{drawn by} solid \edt{lines}. The normalized circulation provides a quantitative measure of the strength of the LEVs, whereas sustainability characterizes their persistence \edt{over the wings} during the downstroke. \edt{These plots are provided for these two metrics calculated} at the root, midspan, and near-tip regions for all three bird\edt{-like} configurations at $\mbox{Re}=10,000$ and all values of $\mbox{St}$ \edt{considered in this work}. \edt{For clarity,} a sustainability value of unity indicates that LEVs remain present within the corresponding spanwise region throughout the complete downstroke, whereas a value below unity indicates that LEVs are \edt{found to be} present only during a fraction of the downstroke. \edt{Hence,} the combined evaluation of these quantities characterizes both the strength and persistence of the LEVs. \edt{Here, Figs.}~\ref{fig:Root_mid_span_circ}\edt{a--\ref{fig:Root_mid_span_circ}c, \ref{fig:Root_mid_span_circ}d--\ref{fig:Root_mid_span_circ}f, and (\ref{fig:Root_mid_span_circ}g--\ref{fig:Root_mid_span_circ}i} correspond to $\mbox{St}=0.18$, $0.225$, and $0.27$, respectively. 

The plots in \edt{Fig.\ref{fig:Root_mid_span_circ}a for} $\mbox{St}=0.18$ \edt{reveal that} $\mbox{Bird}~1$ \edt{has} the \edt{strongest LEV} at the root \edt{of its wing}, followed by the midspan, with \edt{no its measurable presence} near the tip. \edt{On the other sides,} $\mbox{Bird}~2$ and $\mbox{Bird}~3$ exhibit different distributions of these performance indicators here, with \edt{stronger $\mbox{LEVs}$} reaching \edt{their} highest value at the midspans. Despite these differences in \edt{circulations of $\mbox{LEVs}$}, the sustainability remains unity at the root and midspan for all three configurations. \edt{Nonetheless}, the lower sustainability near the tip \edt{of the wings elucidates} that $LEVs$ \edt{could sustain themselves there} only during a limited portion of the downstroke in this region. \edt{One distinguishing factor for $\mbox{Bird}~3$ is the relatively stronger circulation of its $\mbox{LEV}$ near the tip apparent in Fig.~\ref{fig:Root_mid_span_circ}c. For} $\mbox{St}=0.225$, the three configurations exhibit a similar spanwise trend in circulation, with higher values at the root followed by a reduction toward the midspan and near-tip regions. The LEVs remain present throughout the downstroke at the root and midspan, whereas differences among the configurations become evident near the tip. $\mbox{Bird}~2$ and $\mbox{Bird}~3$ exhibit greater near-tip sustainability than $\mbox{Bird}~1$. \edt{Figures~\ref{fig:Root_mid_span_circ}g, \ref{fig:Root_mid_span_circ}f, and \ref{fig:Root_mid_span_circ}h for $\mbox{St}=0.27$ show that} the normalized circulation decreases progressively from the root toward the tip for all three \edt{cases}. $\mbox{Bird}~3$ exhibits the highest normalized circulation at the root, while differences among the configurations also remain evident at the midspan. \edt{We also notice here that an increase in $\mbox{St}$ enhances} the near-tip circulation, particularly for $\mbox{Bird}~2$ and $\mbox{Bird}~3$. The corresponding increase in sustainability \edt{points out} that $\mbox{LEVs}$ remain present over a greater fraction of the downstroke near the tip \edt{for larger $\mbox{St}$}. 



\begin{figure}[H]
    \centering
    \makebox[\textwidth][c]{%
        \includegraphics[width=0.6\textwidth]{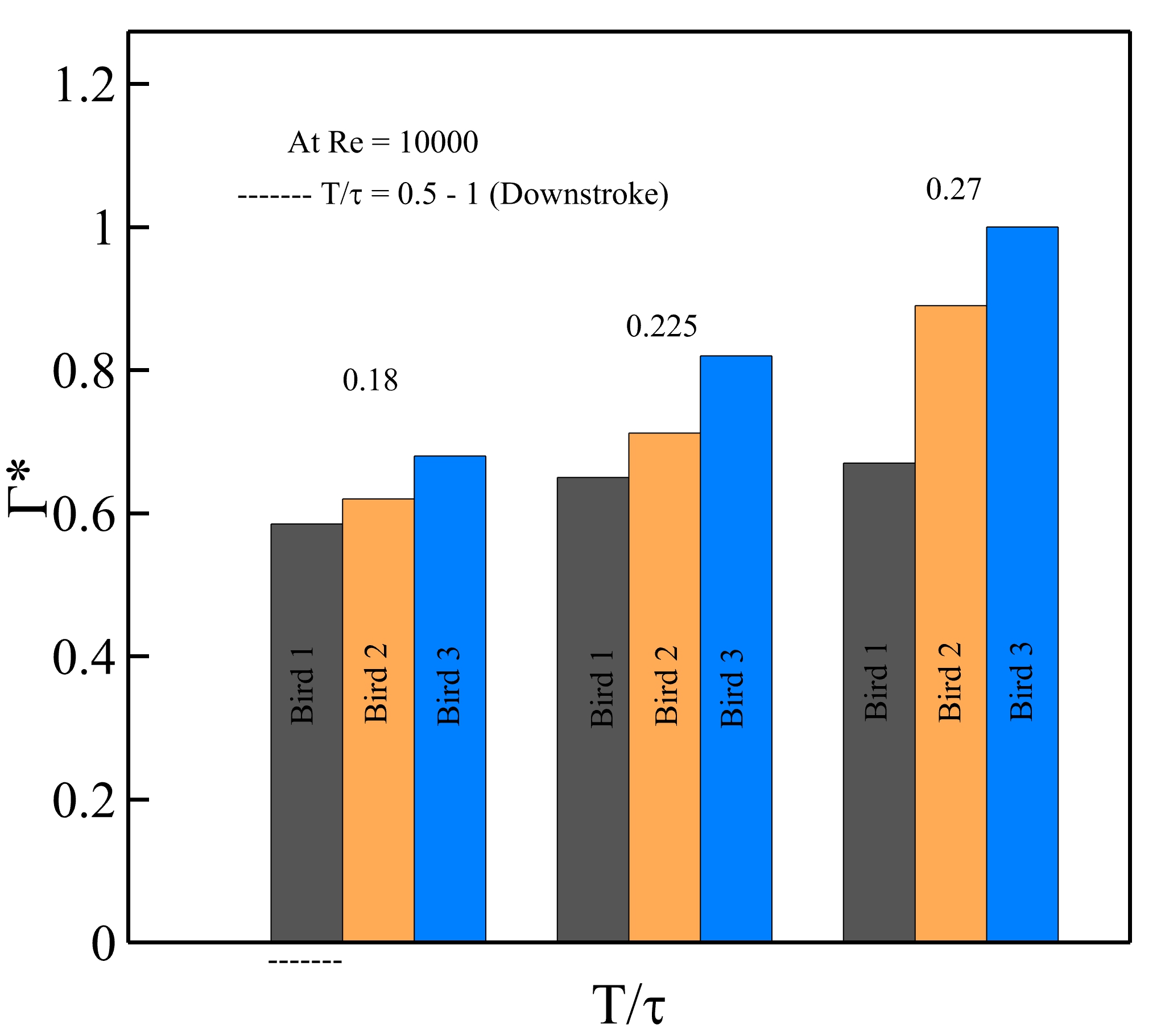}
    }
    \caption{Comparison of the overall normalized LEV circulation during the downstroke for the three bird-wing configurations at $\mbox{Re}=10^4$ and $\mbox{St}=0.18$, $0.225$, and $0.27$}
    \label{fig:Total_LEV_circulation}
\end{figure}

\edt{This quantification} characterize the local behavior of the LEVs, \edt{but}, the spanwise distributions alone do not directly quantify the overall circulation associated with each wing-body configuration. Therefore, the circulation is further evaluated over the wing during the downstroke to provide a \edt{more robust and direct} comparison \edt{between the aerodynamics of} the three configurations. Figure~\ref{fig:Total_LEV_circulation} compares the total normalized circulation \edt{of $\mbox{LEVs}$} during the downstroke \edt{for all cases at $\mbox{Re}=10^4$}. At $\mbox{St}=0.18$, $\mbox{Bird}~1$ exhibits the lowest overall circulation, followed by $\mbox{Bird}~2$, while $\mbox{Bird}~3$ \edt{carries} the highest value. As \edt{$\mbox{St}$} increases to $0.225$ and $0.27$, the overall circulation of \edt{the coherent flow structures} \edt{enhances} for all three configurations, while the differences among them become more pronounced. $\mbox{Bird}~1$ consistently exhibits the lowest circulation, whereas $\mbox{Bird}~3$ maintains the highest circulation across all \edt{kinematic conditions here. Therefor, our qualitative as well as quantitative analyses at different levels provide evidences that distributions of feathers-like morphological features over the span of avian wings and serrations on their trailing sides may substantially help birds modify the spanwise development of \edt{stronger and more sustainable coherent flow features} during downstroke \edt{of their flapping motions.}}


\section{Conclusions}
 
\edt{Our} present \edt{work} investigates the influence of \edt{an avian wing's} geometry and feather morphology on the aerodynamic performance and three-dimensional vortex dynamics of three flapping bird-wing configurations. Bird~1 represents a smooth and continuous geometry \edt{of its wings}, Bird~2 incorporates feather-like structures toward the trailing-edge and wing-tip regions, and Bird~3 additionally incorporates layered feather morphology. The aerodynamic performance is evaluated over the investigated Reynolds and Strouhal numbers through the mean lift coefficient, mean drag coefficient, and lift-to-drag ratio, followed by an analysis of formation evolution \edt{of $\mbox{LEVs}$, their} spanwise development, sustainability, and circulation. \edt{Comparisons of} the aerodynamic forces shows distinct responses \edt{from} the three configurations. Bird~2 generally produces the highest mean drag, while its mean lift remains comparatively high over the investigated conditions. Bird~3 shows variations in its drag characteristics with changes in $\mbox{Re}$ and $\mbox{St}$, while its mean lift remains close to that of Bird~2 and becomes higher at the higher investigated conditions. \edt{Contrarily}, Bird~1 generally produces lower mean lift. The combined effect of lift and drag is reflected in the comparison of $L/D$, where Bird~3 exhibits the most favorable overall aerodynamic performance, demonstrating a better balance between lift generation and drag production for the investigated conditions. \edt{Moreover,} our analysis shows the formation and evolution of multiple LEVs along \edt{a wing's} span for all three configurations. Differences are observed in their size, coherence, spanwise connectivity, attachment, and breakdown during the downstroke, with Bird~3 retaining comparatively more identifiable coherent vortical structures over \edt{its flapping} wing with more flow attached over the whole wing as well. The quantitative analysis further supports these observations, as Bird~3 exhibits the highest overall normalized circulation of $\edt{LEVs}$ for the investigated $\mbox{Re}$. The root and midspan regions generally sustain stronger and more persistent LEVs, whereas increased flow unsteadiness, breakdown of LEVs, and interactions among the LEV, TEV, and TiV become more prominent toward the wing-tip regions. Overall, these \edt{findings demonstrate} that the aerodynamic influence of feathers depends not only on their presence but also on their geometric arrangement. Bird~2 demonstrates that comparatively high lift accompanied by a higher drag penalty does not necessarily result in improved overall aerodynamic performance. \edt{Oppositely}, the layered feather morphology of Bird~3 provides a more favorable balance between lift and drag while maintaining \edt{stronger LEVs} and more coherent vortex structures. These findings demonstrate the importance of feather arrangement in modifying the aerodynamic performance and vortex dynamics of flapping \edt{wings of birds}.



\section*{Acknowledgment}
M.S.U. Khalid acknowledges the funding support from Lakehead University through the startup grant and the Natural Sciences and Engineering Research Council of Canada (NSERC) through the Discovery and Alliance International grant program. D. Thakur is thankful to Lakehead University for the graduate scholarship and for the support through the Ontario Graduate Scholarship. The simulations reported in this work were performed on the supercomputing clusters administered and managed by the Digital Research Alliance of Canada.  

\section*{Data Availability Statement} 
All data generated are available in the article itself. 


\vskip6pt

\bibliography{references}

\end{document}